\documentclass{rspublic}

\usepackage{amsmath}
\usepackage{amsbsy}
\usepackage{amssymb}
\usepackage{amscd}
\usepackage{amsfonts}
\usepackage{graphics}
\usepackage{subfigure}
\usepackage[authoryear]{natbib}
\usepackage[all]{xy}

\def\br{{\mathbf{r}}}

\def\brp{{\mathbf{r}'}}
\def\bR{{\mathbf{R}}}

\begin{document}

\title[Role of the defect-core]{Role of the defect-core in energetics of vacancies}
\author[V. Gavini]{Vikram Gavini}

\affiliation{Department of Mechanical Engineering, University of Michigan, Ann Arbor, USA}

\label{firstpage}

\maketitle

\begin{abstract}{Electronic structure, Defect-core, Quasi-continuum, Vacancies, Aluminum}
Electronic structure calculations at macroscopic scales are employed to investigate the crucial role of a defect-core in the energetics of vacancies in aluminum. We find that vacancy core-energy is significantly influenced by the state of deformation at the vacancy-core, especially volumetric strains. Insights from the core electronic structure and computed displacement fields show that this dependence on volumetric strains is closely related to the changing nature of the core-structure under volumetric deformations. These results are in sharp contrast to mechanics descriptions based on elastic interactions that often consider defect core-energies as an inconsequential constant. Calculations suggest that the variation in core-energies with changing macroscopic deformations is quantitatively more significant than the corresponding variation in relaxation energies associated with elastic fields. Upon studying the influence of various macroscopic deformations, which include volumetric, uniaxial, biaxial and shear deformations, on the formation energies of vacancies, we show that volumetric deformations play a dominant role in governing the energetics of these defects. Further, by plotting formation energies of vacancies and di-vacancies against the volumetric strain corresponding to any macroscopic deformation, we find that all variations in the formation energies collapse on to a \emph{universal curve}. This suggests a universal role of volumetric strains in the energetics of vacancies. Implications of these results in the context of dynamic failure in metals due to spalling are analyzed.
\end{abstract}

\section{Introduction}

A wide range of materials properties are significantly influenced by various defects present in materials---examples include the role of dislocations in plastic deformation, vacancies in creep and spalling, dopants in semiconductor properties, and domain walls in ferroelectric properties. These macroscopic properties of solids, and others, are determined not by a single defect, but by a complex interaction of many defects and external loads over a wide range of time-scales. The behaviour of these defects determined by their energetics, which include nucleation, kinetics, and interaction between various defects, have long been studied through elastic interactions (cf. e.~g. Eshelby 1951, 1956; Zhang \textit{et al.} 2006; Garikipati \textit{et al.} 2006; Phillips 2001 for a comprehensive overview). Though elastic interactions have provided tremendous insights into the properties of defects over the past few decades, one serious limitation of such a treatment is the inability to account for defect-cores. The core of any defect is governed by \emph{quantum-mechanical} interactions at the sub-angstrom length-scale, and determined by the electronic structure of the material. In fact, as it will be demonstrated in this paper, it is the core of a defect that determines the nature and strength of elastic fields produced by a defect. Thus, a complete description of defects must include both elastic interactions at the coarse (micrometer and beyond) scale and also electronic structure of defect-cores at the fine (sub-angstrom) scale.

Traditional implementations of electronic structure calculations have for the most part relied on the use of a plane-wave basis and periodic boundary conditions (cf. e.~g. Martin 2004). Although plane-wave basis has been widely regarded as the most efficient complete basis set for computing the electronic structure of a perfect solid, it is not a desirable choice in the study of defects due to the unrealistic periodic geometries used in these calculations. Moreover, the computational complexity associated with electronic structure calculations has limited these investigations to computational domains (cell-sizes) consisting of a few hundred atoms. The periodicity restrictions, in conjunction with cell-size limitations, limits the scope of these calculations to very high concentrations of defects that rarely---if ever---are realized in nature. Typical concentrations of defects, say vacancies, is about a few parts per million (Fluss 1984). But, contemporary electronic structure calculations have only been able to access cell-sizes corresponding to concentrations of the order of a few thousand parts per million, which are orders of magnitude larger concentrations than those that are realized in nature. Thus, to be make contact with experiments, understand and predict the properties of defects in the \emph{dilute limit} as they exist in nature, electronic structure calculations on multi-million atoms (macroscopic scales) are needed.

This significant hurdle has been overcome by the recent development of quasi-continuum orbital-free density-functional theory (QC-OFDFT) (Gavini \textit{et al.} 2007a,b), thus paving the way for an accurate electronic structure study of defects in materials. This method has enabled for the first time a calculation of the electronic structure, using orbital-free density-functional theory (OFDFT) (cf. e.~g. Wang 1999), of samples with millions of atoms subjected to arbitrary boundary conditions. QC-OFDFT is a multi-scale method that enables systematic coarse-graining in such a manner that resolves detailed information in regions where it is necessary (such as in the immediate vicinity of the defect, resolving the electronic structure of the core) but adaptively samples over details where it is not (such as in regions far away from the defect, accounting for long-ranged elastic and electrostatic effects) without significant loss of accuracy. This is achieved through a real-space formulation of OFDFT, and constructing an adaptive basis set using a finite-element discretization. Importantly, the method is \emph{seamless}, treats the quantum-mechanical interactions and long-ranged elastic effects on an equal footing, uses OFDFT as its only input, and enables convergence studies of its accuracy.

In a recent work, QC-OFDFT was used to study the properties of vacancies in aluminum (Gavini \textit{et al.} 2007b). The formation energy of a mono-vacancy (single vacancy) in aluminum was computed to be 0.72 eV. This is in agreement with other reported density-functional theory calculations (cf. e.~g. {Turner} \textit{et al.} 1997), and experimental estimates predict the value to be 0.66 eV (cf. e.~g. Triftsh\"auser \textit{et al.} 1975). It is striking to note that, of the total formation energy, the contribution from the defect-core (defect core-energy), which is solely governed by the electronic structure, is 0.78 eV. Whereas the contribution from atomic relaxations that produce elastic effects is only 0.06 eV, which is less than 10\% of the total formation energy. These results are again consistent with other density-functional theory calculations (Gillan 1989; Turner \textit{et al.} 1997). Although the contribution from defect-cores to the energetics of defects is very significant, it is often assumed to be an inconsequential constant, and the energetics of defects are often only described through elastic interactions---this is the underlying hypothesis in most mechanics descriptions of defects.

In this work, large-scale electronic structure calculations employing QC-OFDFT are analyzed to investigate the role of defect-cores in energetics of vacancies. In particular, we seek to determine if the defect core-energy is indeed an inconsequential constant. To this end, we compute the energetics of vacancies in aluminum subject to macroscopic volumetric deformations. The existence of a dependence of vacancy formation energies on macroscopic deformations has been initially reported in a Letter (Gavini 2008). Here we present a detailed analysis of the energetics, its implications on the defect-core energy, its relationship to the core electronic structure and displacement fields, the universal role of volumetric strains in the defect energetics, and the new physical insights provided by these results into the mechanism of spalling in metals. These investigations have revealed that contributions from the defect-core---formation energies where the elastic fields generated by the defect-core are suppressed---are strongly influenced by macroscopic volumetric deformations. The dependence is so significant that the core-energy of a mono-vacancy in aluminum varied from 4.73 eV at -0.36 volumetric strain (95 GPa pressure) to 0.07 eV at 0.33 volumetric strain (-21 GPa pressure). Moreover, the nature and strength of displacement fields around the vacancy-core completely changes with imposed volumetric strain. The computed displacement field is radially inward under negative (compressive) volumetric strains as well as under zero strain, but changes to a radially outward nature for large positive (tensile) volumetric strains. This change appears to be closely associated with the shrinking size of the vacancy-core that is observed from the electronic structure. These results suggest that core-energies of vacancies are not a constant, and in fact vary significantly with the state of deformation present at the defect-core. This dependence in the case of vacancies, and possibly other defects, can alter their behaviour, including nucleation and kinetics, over varying times scales.

Besides studying the influence of volumetric deformations, the dependence of mono-vacancy and di-vacancy (vacancy complexes formed from two vacancies) energetics on other macroscopic deformations---uniaxial, biaxial, shear---are also investigated. Both mono-vacancy and di-vacancy formation energies are found to be influenced by uniaxial and biaxial deformations, however there is no significant influence of shear deformations on the formation energies of these defects. This suggests the possibility of volumetric strain associated with a macroscopic deformation playing a dominant role in influencing the formation energies of these defects. By plotting the formation energies of these defects against the volumetric strain associated with any deformation, we determine that volumetric strain is indeed the universal parameter governing the formation energies of both a mono-vacancy and di-vacancies. Noting that the elastic fields generated by vacancies predominantly interact only with volumetric deformations, we conclude that volumetric strain is the \emph{universal parameter} governing the vacancy core-energies.

The presentation of the work is organized as follows. Section \ref{overview} provides a brief overview of quasi-continuum orbital-free density-functional theory and salient features of the method that enable electronic structure calculations at macroscopic scales. We refer to Gavini \textit{et al.} (2007a,b) for a more comprehensive discussion on the method. Section \ref{Vacancies} provides a discussion on electronic structure studies of vacancies. Details of the simulations and a discussion of the results are provided in section \ref{Results}. We conclude in section \ref{Conclusions} providing new physical insights into energetics of vacancies and its implications on spalling in metals.

\section{Overview}\label{overview}
\subsection{Real-space formulation of orbital-free density-functional theory}\label{RealSpace}
Traditionally, most density-functional calculations have been
performed in Fourier-space using plane-wave basis functions
({Martin} 2004). The choice of a plane-wave basis for electronic
structure calculations has been the most popular one, as it lends
itself to a computation of the electrostatic interactions
naturally using Fourier transforms. However, such a Fourier-space formulation has very serious limitations in describing defects in materials. Firstly, it requires periodic boundary conditions, thus limiting an investigation to a periodic array of defects. This periodicity restriction in conjunction with the cell-size limitations ($\sim 200$ atoms) arising from the enormous computational cost associated with electronic structure calculations, limits the scope of these studies to very high concentrations of defects that are not realized in nature. Importantly, plane-wave basis functions used in Fourier-space formulations provide a uniform spatial resolution, which is not desired in the description of defects in materials. Often, higher resolution is required in the description of the core of a defect and a coarser resolution suffices away from the defect-core. This in turn makes Fourier-space formulations \emph{computationally inefficient} in the study of defects in materials. Further, from a numerical viewpoint, plane-wave basis functions are non-local in the real-space, thus resulting in a dense matrix which limits the effectiveness of iterative solutions. Also, a plane-wave basis requires the evaluation of Fourier transforms which affect the scalability of parallel computation.

For all the above reasons, and since it is a key-component in the coarse-graining of electronic structure calculations, a real-space formulation for orbital-free density-functional theory is proposed. The ground-state energy in density-functional theory is given by
(cf. e.~g. Finnis 2003; Parr \& Yang 1989)
\begin{eqnarray}\label{Energy}
    E(u, \bR)
    =
    T_s(u)+E_{xc}(u)+E_H(u)+E_{ext}(u,
    \bR)+E_{zz}(\bR)\,,
\end{eqnarray}
where $u(\br) = \sqrt{\rho(\br)}$ is the ground-state square-root electron-density; $\bR = \{\bR_1, \dots,
\bR_M\}$ is the collection of nuclear positions in the system;
$T_s$ is the kinetic energy of non-interacting electrons; $E_{xc}$
denotes the exchange correlation energy; $E_H$ is the classical
electrostatic interaction energy between electrons, also referred
to as Hartree energy; $E_{ext}$ is the interaction energy of
electrons with external field induced by nuclear charges; and
$E_{zz}$ denotes the repulsive energy between nuclei.

Orbital-free density-functional theory (OFDFT) is a version of density-functional theory where the kinetic energy of non-interacting electrons, $T_s$, is modeled, as opposed to the standard Kohn-Sham density-functional theory (KSDFT), where this is evaluated exactly within the mean-field approximation by solving an effective single electron Schr$\ddot{o}$dinger's equation. In material systems whose electronic structure is close to a free-electron gas, e.g. simple metals, aluminum etc., very good orbital-free models for the kinetic energy term are available which have been shown to accurately predict a wide range of properties in these materials. A simple choice for this is
the Thomas-Fermi-Weizsacker (TFW) family of functionals (Parr \& Yang 1989),
which have the form
\begin{eqnarray}\label{E:9}
    {T}_{s}(u)
    =
    C_F\int{{u}^{10/3}(\br)d\br}
    +
    \frac{\lambda}{2}
    \int{{{|\nabla u(\br)|}^2}
    d\br},
\end{eqnarray}
where $C_F=\frac{3}{10}{(3\pi^{2})}^{2/3}$, and $\lambda$ is a parameter. More recently, there have
been efforts (Wang \& Teter 1992; Samrgiassi \& Madden 1994; Wang \textit{et al.} 1998, 1999) to improve these orbital-free kinetic energy functionals by
introducing an additional non-local term called the kernel energy.
These kinetic energy functionals have a functional form given by
\begin{eqnarray}\label{E:10}
{T}_{s}(u)=C_F\int{{u}^{10/3}(\textbf{r})d\textbf{r}}+\frac{1}{2}\int{{|\nabla u(\textbf{r})|}^2}
+\int{\int{f(u(\textbf{r}))K(|\textbf{r}-\textbf{r}^{'}|)f(u(\textbf{r}^{'}))d\textbf{r}d\textbf{r}^{'}}},\notag\\
\end{eqnarray}
where $f$ and $K$ are chosen such that $T_s(u)$ satisfies the linear response of homogeneous non-interacting electron gas which is known explicitly (cf. e.~g. Wang \& Teter 1992).

$E_{xc}$ denotes the exchange correlation energy which describes the quantum-mechanical interactions for which accurate models for most systems are available. The Local Density Approximation (LDA) (Ceperley \& Alder 1980; Perdew \& Zunger 1981) given by
\begin{eqnarray}
E_{xc}(u)=\int{\epsilon_{xc}(u^2(\br))u^2(\br)
d\br},
\end{eqnarray}
where $\epsilon_{xc}$ has a parameterized form, has been shown to capture the exchange and correlation effects for most systems accurately.

The last three terms in energy functional (\ref{Energy}) are electrostatic and are given by:
\begin{eqnarray}
E_H(u) &=& \frac{1}{2}\int\int
    {\frac{u^2(\br)u^2(\brp )}{|\br -\brp |}d\br d\brp }, \\
E_{ext}(u,\bR) &=& \sum_{I=1}^{M} \int{u^2(\br)\frac{Z_I}{|\br-\bR_I|}d\br}, \\
E_{zz}(\bR) &=&  \frac{1}{2}
    \sum_{I=1}^{M}
    \sum_{\substack{J=1\\J\neq{I}}}^{M}
    \frac{{Z}_{I}{Z}_{J}}{|\bR_{I}-\bR_{J}|},
\end{eqnarray}
where $Z_I$ denotes the charge of the nucleus located at $\bR_I$, $I=1, 2,\ldots, M$.
The energy functional given by equation (\ref{Energy}) is local in real-space except for
the electrostatic interaction energy, and some forms of the kinetic energy functionals, e.g. those given by equation (\ref{E:10}). For this reason, evaluation of these energy terms is the most computationally intensive part of the calculation of the energy functional.
However, noticing that $\frac{1}{|\br-\brp|}$ kernel is the Green's function of the Poisson's equation, the electrostatic terms can be expressed locally as the following \emph{variational} problem,
\begin{equation}\label{E:17}
\begin{split}
    &
    \frac{1}{2} \int\
    \int{\frac{u^2(\br)u^2(\brp )}
    {|\br -\brp |}d\br d\brp }
    +
    \int\int{\frac{u^2(\br)b(\brp; \bR )}
    {|\br -\brp |}d\br d\brp }
    +
   \frac{1}{2}\int\int
    {\frac{b(\br; \bR)b(\brp; \bR )}{|\br -\brp |}d\br d\brp }
    \\ &=
    - \inf_{\phi\in H^1_0(\mathbb{R}^3)}
    \left\{
    \frac{1}{8\pi}
    \int{|\nabla\phi(\br)|^{2}d\br}
    -
    \int{(u^2(\br)+b(\br; \bR))\phi(\br)d\br}
        \right\}\,,
\end{split}
\end{equation}
where $\phi$ denotes a trial function for the electrostatic potential of the system of charges, and $b(\textbf{r};\bR)$ denotes the regularized nuclear charges corresponding to the pseudopotentials that provide an external potential for valence electrons. We remark that the left hand side of equation (\ref{E:17}) differs from the sum of electrostatic terms by the self energy of the nuclei, which is an inconsequential constant.

Turning to the non-local kinetic energy terms (kernel energies) given by equation (\ref{E:10}), the approach suggested by Choly \& Kaxiras (2002) is used to approximate the kernel in the reciprocal space by a rational function. Under this approximation, whose error can be systematically controlled, we find that the kernel energies have a local form given by (Gavini \textit{et al.} 2007a)
\begin{subequations}\label{kernelmin}
\begin{align}
T_s(u)=C_F\int{{u}^{10/3}(\br)d\br}
    +
    \frac{1}{2}
    \int{{|\nabla u(\br)|}^2}
    +
    \sum_{j=1}^{m}\frac{2}{C_j}Z_j(u)+(\sum_{j=1}^{m}P_j)\int{f^2(u(\br))d\br}\\
Z_j(u)=\inf_{w_j\in{X_w}}\left\{{\frac{C}{2}}\int|\nabla{w_j}(\br)|^2d\br+\frac{Q_j}{2}\int{w_j^2(\br)d\br}+C_j\int{w_j(\br)f(u(\br))}d\br\right\}\,\,\,\,\,\,\,\,j=1\dots m\,,
\end{align}
\end{subequations}
where $C$, $C_j$, $Q_j$ and $P_j$, $j=1,2,\ldots m$, are constants
determined from a fitted rational function with degree $2m$, and $X_w$ is a suitable function space. We refer to the minimizers of the variational problem in
equation (\ref{kernelmin}b) as \emph{kernel potentials}.

Finally, the problem of determining the ground-state energy, ground-state electron-density and
the equilibrium positions of the nuclei can be expressed as the
minimum problem
\begin{subequations}\label{eq:Inf}
\begin{align}
    \inf_{\bR \in \mathbb{R}^{3M}}\inf_{u\in{X_{u}}}
    E(u,\bR)& &\\
   \text{subject to: }  \int{u^2(\br)d\br} = N, & &
\end{align}
\end{subequations}
where $X_{u}$ is a suitable function space, and $N$ denotes the total number of electrons in the system. A full account of the formulation, suitable function spaces, and the well-posedness of the variational formulation may be found in Gavini \textit{et al.} (2007a).

The ground state energy of any materials system as formulated in equation (\ref{eq:Inf}) is local and has a variational structure. A finite-element basis set (Brenner \& Scott 2002) which respects this mathematical structure and allows for arbitrary boundary conditions and complex geometries is a natural choice to discretize and compute. Moreover, the compact support of a finite-element basis is a desirable property in an implementation of the formulation on parallel computing architecture, which derives considerable importance owing to the computational complexity associated with electronic structure calculations. If $T_h$ denotes the finite-element triangulation, and $X^h_u$ denotes the subspace spanned by the corresponding finite-element basis functions that becomes increasingly dense in $X_u$, the variational problem given by equation (\ref{eq:Inf}) reduces to a constrained minimization problem given by,
\begin{subequations}\label{eq:Inf2}
\begin{align}
    \inf_{\bR \in \mathbb{R}^{3M}}\inf_{u^h\in{X^h_{u}}}
    E(u^h,\bR)& &\\
   \text{subject to: }  \int{({u^h}(\br))^2d\br} = N. & &
\end{align}
\end{subequations}
Convergence of the finite-element approximation was rigorously proved in a previous work (Gavini \textit{et al.} 2007a) through $\Gamma-$convergence, which is a variational notion of convergence for non-linear functionals (Dal~Maso 1993). The efficacy of using a finite-element basis set in computing the electronic structure via equation (\ref{eq:Inf2}) has been studied by computing the ground-state properties of aluminum clusters, where clusters as large as 3700 atoms were considered (Gavini \textit{et al.} 2007a).

The use of a finite-element basis set in the computation of the electronic structure, which is amenable to adaptive coarse-graining, is a significant departure from previous numerical methods relying on plane-wave basis sets that have a uniform resolution in real-space. There have been previous efforts which employed a finite-element basis set in electronic structure calculations (Pask \textit{et al.} 1999; Pask \& Sterne 2005). However, these efforts did not exploit the adaptive nature of the basis set, which is the key feature that is exploited in the construction of a quasi-continuum reduction described in the following section---one that will enable electronic structure calculations at macroscopic scales.

\subsection{Quasi-continuum reduction}

The real-space formulation of orbital-free density-functional theory and its finite-element discretization as proposed in section 2(\ref{RealSpace}) is attractive, as it gives \emph{freedom from periodicity}, which is important in modelling defects in materials. But, the complexity of these calculations limit investigations to systems consisting of at most a few thousand atoms. On the other hand, materials properties are influenced by defects---vacancies, dopants, dislocations, cracks, free surfaces---in small concentrations (parts per million). An accurate understanding of such defects must not only include the electronic structure of the core of the defect, but also the elastic and electrostatic effects on the macro-scale. This in turn requires calculations involving millions of atoms, well beyond the current capability.%

This significant challenge is overcome by a \emph{Quasi-Continuum reduction} of the proposed real-space finite-element formulation, and enables computation of the electronic structure of  multi-million atom systems without significant loss of accuracy. This is a multi-scale scheme which facilitates a systematic \emph{coarse-graining} of electronic structure calculations in a \emph{seamless} manner that resolves detailed information in regions where it is necessary (such as in the immediate vicinity of the defect) but adaptively samples over details where it is not (such as in regions far away from the defect) without significant loss of accuracy. The real-space formulation of orbital-free density-functional theory, and a finite-element discretization which is amenable to coarse-graining are crucial steps in its development. The approach is similar in spirit to the quasi-continuum (QC) method developed in the context of interatomic potentials (cf.
e.~g., Tadmor \textit{et al.} 1996; Knap \& Ortiz 2001) to coarse-grain the displacement field of an atomic lattice. However, the proposed approach addresses a more challenging problem to also coarse-grain the \emph{electronic fields}---electron-density, electrostatic potential, and kernel potentials---that exhibit sub-atomic oscillations.%

\begin{figure}
\centering
    {\scalebox{0.7}{\includegraphics{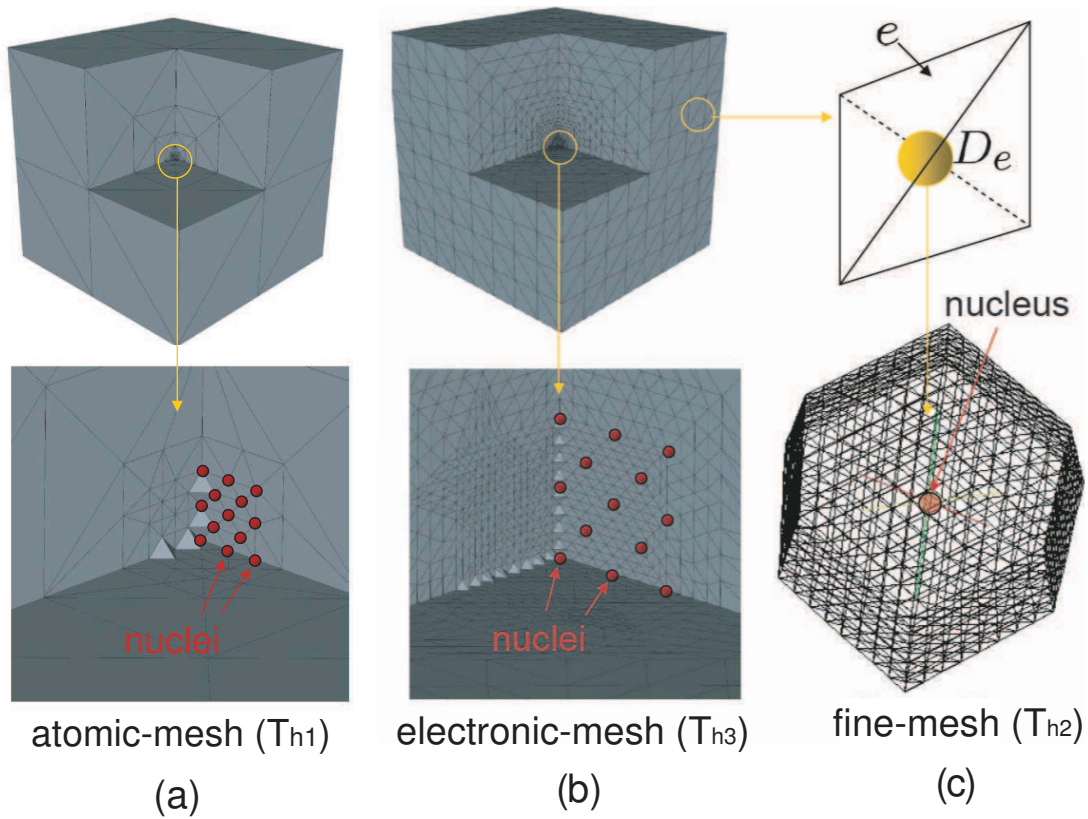}}}
    \caption{\label{Multigrid}Hierarchy of triangulations used in the QC reduction. (a) Atomic-mesh ($T_{h_1}$) used to interpolate
nuclei positions away from the fully-resolved defect-core; (b)
Electronic-mesh ($T_{h_3}$) used to represent the corrector fields. It
has subatomic resolution in the defect-core, and coarsens away
from the defect-core. (c) Fine auxiliary mesh ($T_{h_2}$) is used to sample the Cauchy-Born predictor fields within an integration lattice unit cell, $D_{e}$, in each element $e$.\newline}
\end{figure}

The independent unknowns in an electronic structure calculation using orbital-free density-functional theory are the nuclear positions, and the electronic fields comprising of electron-density, electrostatic potential, and kernel potentials. In the proposed quasi-continuum reduction of electronic structure calculations, as in the conventional QC approach, the nuclei positions are
interpolated from the positions of representative nuclei through a triangulation $T_{h_1}$,
Figure~\ref{Multigrid}(a). Near the defect-core, all nuclei are
represented, whereas away from the defect-core the interpolation becomes coarser and a
small fraction of the nuclei determine the displacements of the
rest. We refer to this computational mesh as the
\emph{atomic-mesh}. Turning to the electronic fields, they exhibit a fine-scale structure on the sub-atomic length-scale. An accurate resolution of these fields on the sub-angstrom length-scale is necessary to achieve chemical accuracy in the computed ground-state properties, which presents a unique challenge both in terms of the computational cost and memory requirements. However, in regions far away from a defect, where the displacement fields are slowly varying, formal mathematical results (Blanc \textit{et al.} 2002) have shown that the electronic structure is given by a local periodic calculation using the Cauchy-Born deformation of the underlying lattice. Thus, to realize the coarse-graining of the electronic structure, two unstructured triangulations of the domain are introduced as shown in Figure~\ref{Multigrid} (b)-(c) to provide a complete description of the discrete electronic fields:

\noindent (i) A triangulation $T_{h_3}$, subatomic close to lattice defects and increasingly coarser away from the defects, which is labeled as the \emph{electronic-mesh}.\newline
(ii) An auxiliary subatomic triangulation $T_{h_2}$ that resolves a lattice unit-cell to capture the subatomic oscillations in the electronic fields, which is labeled as the \emph{fine-mesh}.

For convenience, the triangulations are restricted in such a way that $T_{h_3}$ is a
sub-grid of $T_{h_1}$. $X_{h_1}$, $X_{h_2}$ and $X_{h_3}$ are labeled as the corresponding finite-element approximation spaces. The representation of electronic fields, comprising of square-root electron-density, electrostatic potential and kernel potentials, is decomposed as
\begin{equation}
{u^h} = {u_0^h}+{u_c^h}, \quad \phi^h = \phi_0^h+\phi_c^h,\quad  w^h_j={w^h_j}_0+{w^h_j}_c\,\qquad j=1,\ldots m, %
\end{equation}
where ${u_0^h},\phi_0^h,{w^h_j}_0\in X_{h_2}$ denote the
predictor for electronic fields and is obtained through a local periodic calculation corresponding to the Cauchy-Born deformation of every element of ${T_{h_1}}$. ${u_c^h},\phi_c^h,{w^h_j}_c\in X_{h_3}$ represents the corrections to be solved for. The predictor for the electronic fields is expected to be accurate away from defect-cores, in regions where the deformation field is slowly varying. Hence, the corrections may be accurately represented by means of a finite-element triangulation such as $T_{h_3}$, that has subatomic resolution close to the defect and coarsens away from the defect to become superatomic.%

The minimization problem given by equation (\ref{eq:Inf2}) now reduces to a minimization over the corrector fields, which are resolved on a coarse-grained mesh with far fewer degrees of freedom compared to the full problem,%
\begin{subequations}\label{eq:Inf3}
\begin{align}
\inf_{\bR \in X_{h_1}}\inf_{{u_c^h} \in X_{h_3}}
    E({u_0^h}+{u_c^h,\bR)}&&\\
    \text{subject to: } \int({u_0^h}+{u_c^h})^2 d\br=N\,.&&%
\end{align}
\end{subequations}
We note that the auxiliary fine-mesh, $T_{h_2}$, is introduced for purposes of representing the Cauchy-Born predictor and does not introduce any degrees of freedom into the calculation.
The degrees of freedom of the QC-OFDFT analysis are the corrector fields on the electronic-mesh and the positions of representative atoms on the atomic-mesh which are computed from the variational formulation. These fields completely describe the electronic structure and the relaxed atomic structure of the material system.

The convergence properties of the proposed quasi-continuum reduction have been analyzed in a recent work (Gavini \textit{et al.} 2007b), and the significant computational savings afforded by the proposed scheme has made electronic structure calculations at macroscopic scales accessible. Importantly, the proposed scheme treats the quantum-mechanical interactions and the long-ranged elastic interactions on an equal footing by using a single electronic structure theory as the sole input physics, and has paved the way for an accurate electronic structure study of defects in materials.

\section{Energetics of vacancies}\label{Vacancies}

Vacancies in materials are known to play a crucial role in the nucleation, evolution, and kinetics of larger defects, which eventually govern, over varying time-scales, the macroscopic deformation and failure mechanisms observed in a variety of metals. For example, vacancies play a crucial role in the phenomena of creep and ageing in metals that occur over longer time-scales of a few years (Somoza \textit{et al.} 2007). On the other hand, vacancies mediate the failure of metals under shock loading due to spalling, which occurs over very short time-scales of less than a second (Meyers \& Aimone 1983). Further, recent theoretical studies predict that vacancies lubricate dislocation motion (Lu \& Kaxiras 2002), which explain the observed softening in cold-worked high-purity aluminum at very low temperatures. Other examples include the role of vacancies in hydrogen embrittlement of metals (Lu \& Kaxiras 2005), and nucleation of prismatic dislocation loops (Gavini \textit{et al.} 2007c) that are responsible for the hardening phenomenon observed in metals subjected to irradiation.

Given the crucial role of vacancies, many efforts in the past focused on studying the behaviour of these defects using continuum theories, atomistic simulations using empirical potentials, and electronic structure calculations (cf. e.~g. {Phillips} 2001 for a comprehensive overview). Continuum descriptions of defects date back to Eshelby's studies on the interaction of a defect with macroscopic elastic fields that are either generated by external loads or other defects (Eshelby 1951, 1956). Though linear elastic continuum descriptions have given immense insights into the behaviour of defects over the past decades, they suffer from well known drawbacks. Firstly, they admit singular displacement fields and cannot capture the nonlinearities at the defect-core, and hence cannot describe the core of a defect. Secondly, they predict zero interaction between isotropic defects in isotropic media---such as interaction between two vacancies. However, it is well known from experiments and other atomistic calculations that vacancies exhibit non-trivial interactions. Higher order strain-gradient elasticity theories and atomistic calculations using empirical potentials are devoid of the above mentioned deficiencies of linear elastic continuum descriptions (Acharya \& Bassani 2000; Zhang \textit{et al.} 2006; Knap \& Ortiz 2001). However, the various parameters in these theories are usually fitted to bulk properties computed from electronic structure calculations, and the transferability of these models in accurately predicting the behaviour of defects is yet to be ascertained.

The nature of the core of any defect is determined by the complex quantum-mechanical interactions at the sub-angstrom length-scale and can only be determined accurately by electronic structure calculations. There have been many studies in the past to determine the electronic structure of vacancies, especially using density-functional theory (Chetty \textit{et al.} 1995; Turner \textit{et al.} 1997; Wang \textit{et al.} 1999; Carling \textit{et al.} 2000; Uesugi \textit{et al.} 2003). However, most electronic structure calculations performed thus far are restrictive. Owing to the computational complexity of electronic structure calculations, simulations were performed on small cell-sizes ($\sim$ 100 atoms) using periodic boundary conditions. The assumption of periodicity along with cell-size restrictions limits the scope of these calculations to very high concentrations of vacancies that are unrealistic. Further, the small cell-sizes used in these calculations may introduce spurious interactions arising from the long-ranged elastic and electrostatic fields in the presence of defects. Thus, on one hand, continuum theories accurately account for the long-ranged elastic effects in the presence of defects, but fail to describe the core of a defect. On the other, electronic structure calculations can accurately describe the core of a defect, but fail to describe these long-ranged effects due to cell-size restrictions.

Quasi-continuum orbital-free density-functional theory has provided a way to overcome this significant challenge, where the quantum-mechanical interactions at the defect-core, as well as, the long-ranged elastic effects are treated on an equal footing by using a single electronic structure theory. In a recent work, QC-OFDFT was used to study the energetics of vacancies in an undeformed aluminum crystal (Gavini \textit{et al.} 2007b). Computed mono-vacancy formation energies as well as di-vacancies binding energies were in agreement with experimental interpretations. Importantly, these calculations indicate a very strong cell-size (concentration) dependence owing to the long-ranged elastic and electrostatic effects. Specifically, it was found that $\langle110\rangle$ di-vacancies were repulsive for small cell-sizes($<$100 atoms), which is in agreement with previous DFT calculations (Carling \textit{et al.} 2000; Uesugi \textit{et al.} 2003). The same di-vacancies were attractive for larger cell-sizes ($>$1000 atoms) corresponding to realistic vacancy concentrations (Fluss \textit{et al.} 1984) of few parts per million, with binding energies in agreement with experimental interpretations (Ehrhart \textit{et al.} 1991; Hehenkamp 1994). Similar cell-size effects were also observed in larger vacancy clusters where the interaction between various di-vacancies was investigated (Gavini \textit{et al.} 2007c). It was observed that the di-vacancy interaction changes from repulsion to attraction at very large cell-sizes of over 10,000 atoms. This suggests that previous discrepancies between computations and experiments may be the result of small cell-sizes used in computations, and highlights the importance of accounting for the long-ranged elastic and electrostatic effects in the studies on defects.

The studies on the energetics of vacancies using QC-OFDFT have revealed some interesting facts (Gavini \textit{et al.} 2007b). In one of the studies, the formation energy of a mono-vacancy was computed under two scenarios: one where the atomic relaxations are suppressed by restricting the atoms to their bulk positions, and another where the atoms are allowed to relax. The study was aimed at estimating the contribution from the defect-core to the total vacancy formation energy. Upon suppressing the atomic relaxations, the computed energy of the vacancy is solely due to the electronic structure of the core and represents a significant part of the defect core-energy. It is surprising to note that, of the total formation energy of a mono-vacancy in aluminum which is computed to be 0.72 eV, the contribution from the defect-core, solely, is 0.78 eV, and atomic relaxations that generate elastic fields contribute only 0.06 eV. Thus, the elastic contribution is less that $10\%$ of the total formation energy. This estimate is consistent with other electronic structure calculations using density-functional theory (cf. e.~g. Gillan 1989; Turner \textit{et al.} 1997). Though the defect-core has a significant contribution to the total energy of defects, the core-energy is often considered an inconsequential constant and the energetics of defects in most mechanics descriptions are determined only through the elastic fields generated by these defects. In many scenarios, where defects interact with one another or with elastic fields generated by external loading, the state of local deformation at the defect-core can be significantly different compared to an undeformed lattice, which can potentially change the core-energy. To the best of our knowledge, there have not been any studies to date to ascertain whether the defect core-energy is indeed a constant, or if it can change with the local state of deformation at the core, and if so, if it is significant enough to alter the energetics of vacancies.

In this work, electronic structure calculations are used to address this key question, as well as understand how the energetics of vacancies---formation energies and interaction energies---are influenced by macroscopic deformations arising due to externally applied loads and in the presence of extended defects like cracks, surfaces, interfaces or dislocations.  In the light of strong cell-size effects on the properties of vacancies demonstrated in recent studies (Gavini \textit{et al.} 2007b,c), QC-OFDFT method is used in the present study which is free of any cell-size restrictions. Though the limitations of the kinetic energy functionals in OFDFT are well understood, OFDFT has been shown to be a good electronic structure theory to compute materials properties in aluminum (Wang \textit{et al.} 1999). Further, the combination of Thomas-Fermi-Weizsacker kinetic energy functional, local density approximation for the exchange-correlation effects (Perdew \& Zunger 1981) and a modified form of Heine-Abarenkov pseudopotential (Goodwin \textit{et al.} 1990) for aluminum has been demonstrated to correctly predict the bulk and vacancy properties in aluminum (Gavini \textit{et al.} 2007a,b), and is employed in the present study.

\section{Results and Discussion}\label{Results}

The effect of macroscopic deformations on vacancy properties is investigated in terms of the influence of a homogeneous macroscopic strain tensor on the energetics of vacancies. In this discussion, we refer to the strain tensor as the symmetric gradient of a displacement field, $\epsilon_{ij}=\frac{1}{2}(\frac{\partial u_i(\textbf{X})}{\partial X_j}+\frac{\partial u_j(\textbf{X})}{\partial X_i})$, where $u_i(\textbf{X})$ is the displacement of the deformed configuration with respect to the undeformed reference configuration $\textbf{X}$. We note that other measures of strain are widely used in materials undergoing large deformations, but, since a constitutive law is not being attempted in this study, the aforementioned measure of strain is as good as any other measure. Without loss of generality, the coordinate axes 1, 2, and 3 are chosen along [100], [010], and [001] respectively. To compute the energetics of vacancies at a prescribed macroscopic strain, we begin by homogeneously deforming a perfect, undeformed crystal (computational cell) into a state corresponding to the imposed macroscopic strain. We then introduce vacancies into this deformed crystal and compute the formation energies of a mono-vacancy, $\left<110\right>$ di-vacancies, and $\left<100\right>$ di-vacancies. The formation energy of a mono-vacancy at a prescribed macroscopic strain is given by (cf. e.~g. Finnis 2003),
\begin{eqnarray}
    E^{f}_{v}(\epsilon_{ij})
    =
    E(N-1,1;\epsilon_{ij})-\frac{N-1}{N}E(N,0;\epsilon_{ij})\,\,\,,
\end{eqnarray}
where $E(N-1,1;\epsilon_{ij})$ is the energy of a system with $N-1$ atoms and $1$ vacancy under a macroscopic strain $\epsilon_{ij}$, and $E(N,0;\epsilon_{ij})$ is the energy of a system with $N$ atoms and no vacancy under the same strain. The formation energies of $\left<110\right>$ di-vacancies ($E^{f}_{2v\left<110\right>}$) and $\left<100\right>$ di-vacancies ($E^{f}_{2v\left<100\right>}$) are similarly defined, which are vacancy complexes formed from two adjacent vacancies along the $\left<110\right>$ and $\left<100\right>$ crystallographic directions. Homogeneous Dirichlet boundary conditions are imposed on the corrector electronic fields as well as the displacement fields, which imply that the perturbations in electronic and displacement fields in the presence of these defects vanish on the boundaries of the sample. All simulations are performed on large computational domains consisting of about $10^6$ atoms, which are free of any cell-size effects and also correspond to a realistic vacancy concentration in materials of a few parts per million (Fluss \textit{et al.} 1984). Mesh parameters were chosen to keep the numerical error in the formation energy due to discretization and coarse-graining to be within 0.02 eV.

\subsection{Volumetric deformations}

\begin{figure}%
\centering%
\subfigure[]{
{\scalebox{0.6}{\includegraphics{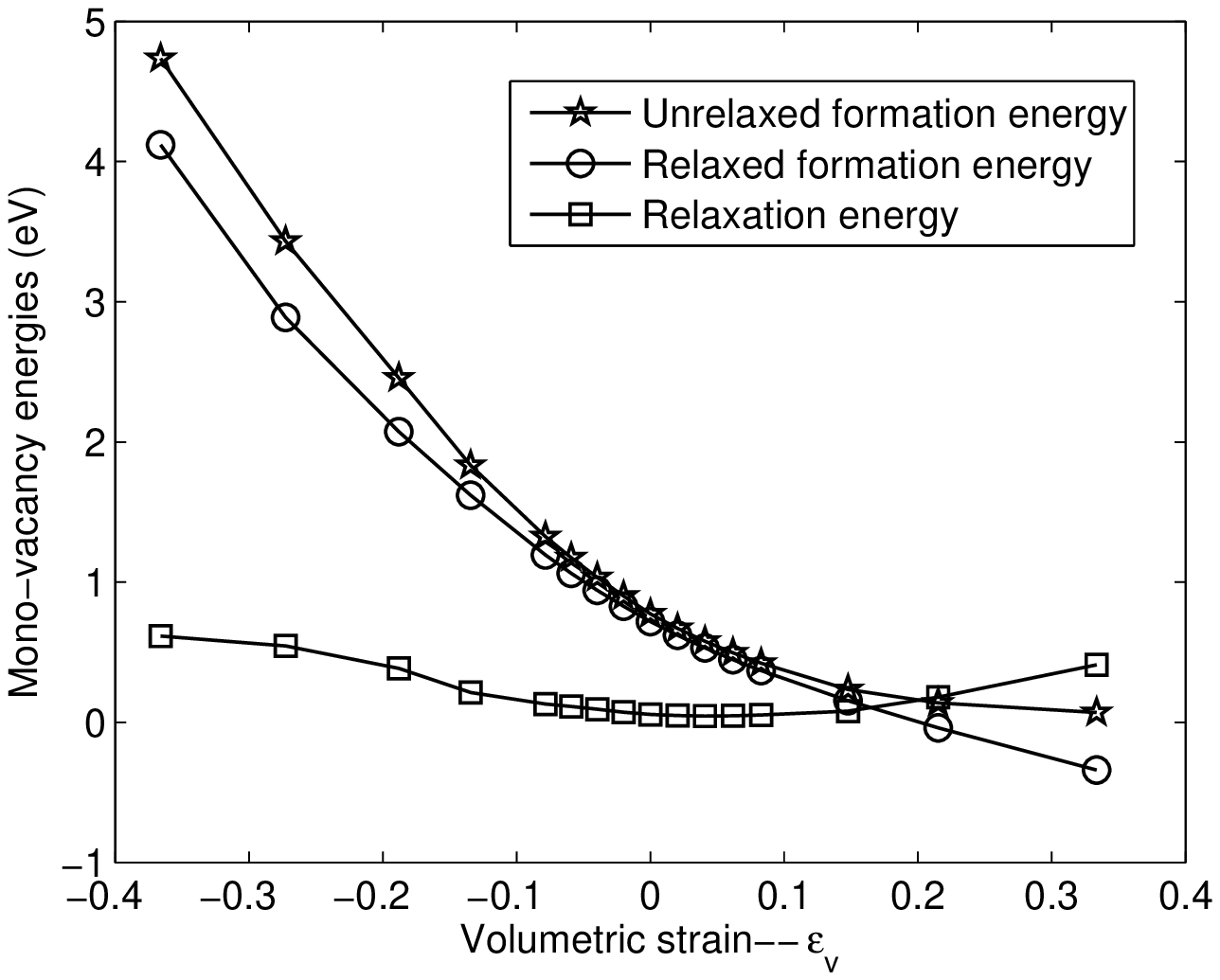}}}}%
\subfigure[]{
{\scalebox{0.6}{\includegraphics{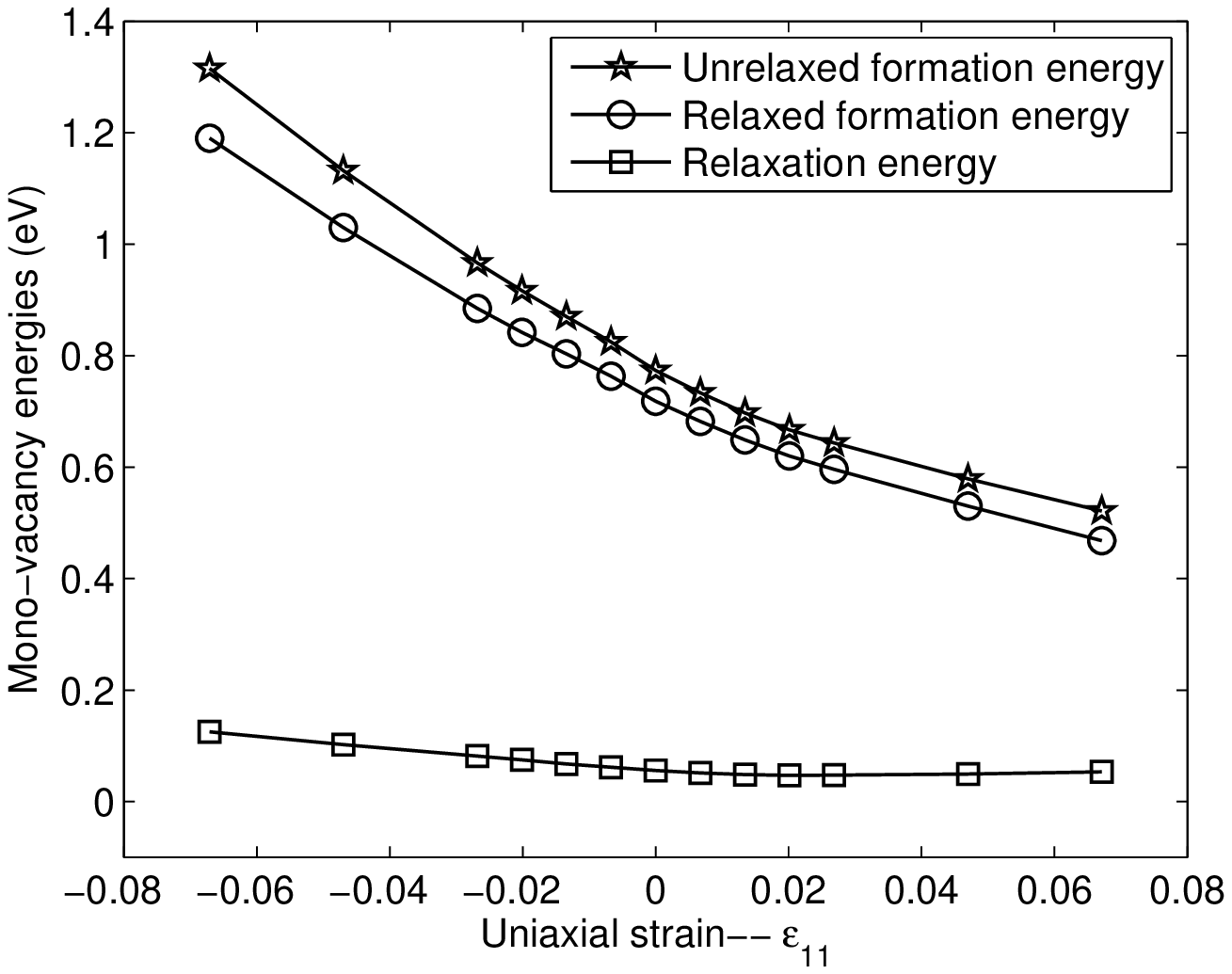}}}}\\%
\subfigure[]{
{\scalebox{0.6}{\includegraphics{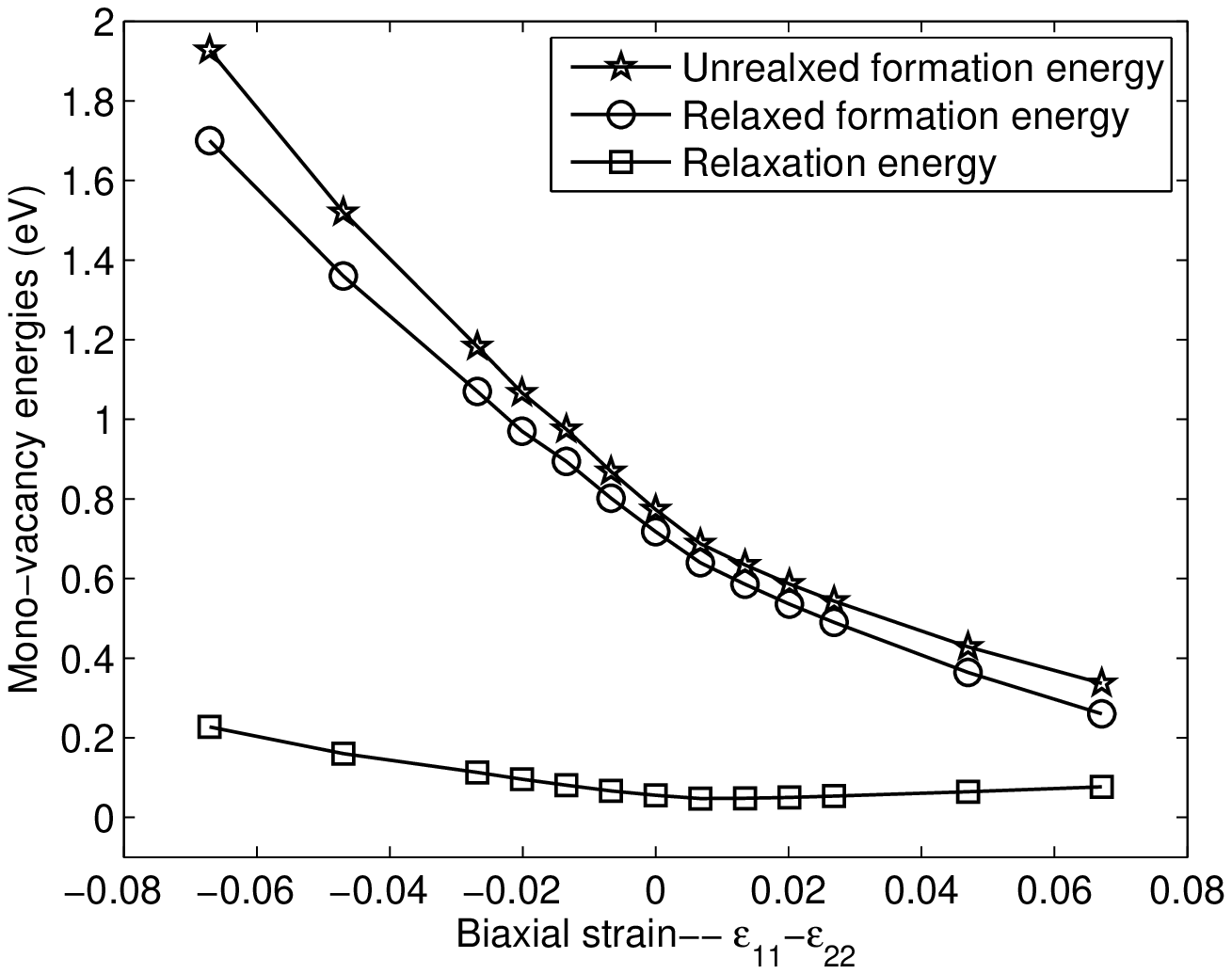}}}}%
\subfigure[]{
{\scalebox{0.6}{\includegraphics{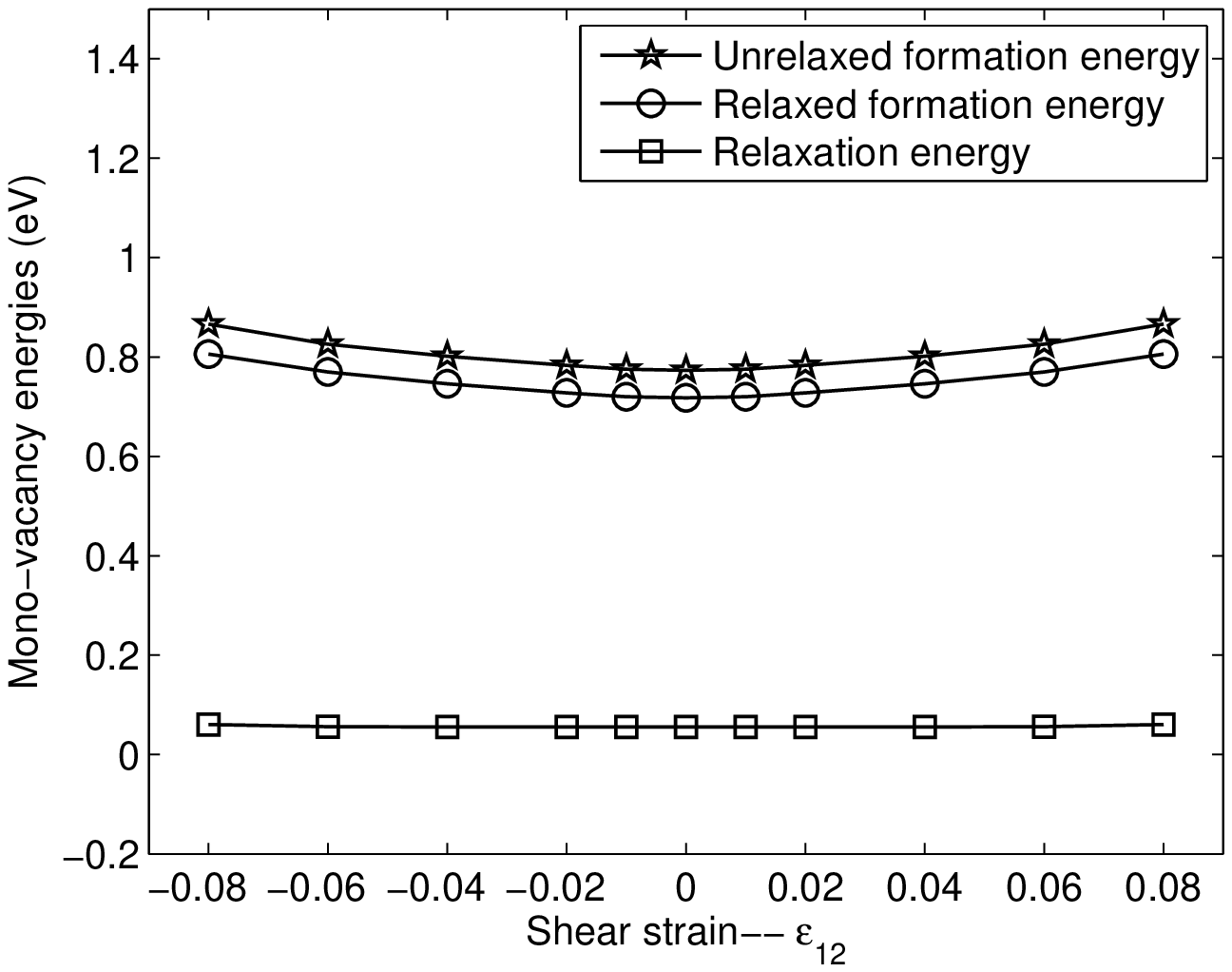}}}}%
\caption{\label{vacancy_deformation} Influence of various macroscopic deformations (strains) on the formation energy of a mono-vacancy: (a) Volumetric strain; (b) Uniaxial strains; (c) Biaxial strains; (d) Shear strains.}%
\end{figure}

In order to investigate the role of defect-cores in energetics of vacancies, we begin by studying the influence of isotropic volumetric deformations on the mono-vacancy formation energy. Any dependence of the formation energy on macroscopic deformations has two possible components: (i) the elastic interaction between the displacement field generated by a vacancy and the imposed macroscopic deformation; (ii) the dependence of the defect core-energy on the macroscopic deformation. To understand the contribution of these two components to the formation energy, we compute the energetics of a mono-vacancy under two scenarios: first by restricting the positions of the atoms to their bulk values determined by the imposed macroscopic strain and suppressing the atomic relaxations arising due to the vacancy (unrelaxed case), and second by allowing for atomic relaxations (relaxed case). In the first case, where the atomic relaxations are suppressed, the contribution to the formation energy is solely from the electronic structure which is often disregarded in a mechanics description as an inconsequential defect core-energy. The influence of isotropic volumetric deformations on the unrelaxed and relaxed formation energies is shown in figure \ref{vacancy_deformation}(a). It is evident from these results that, even without atomic relaxations, there is a significant influence of volumetric deformations on the mono-vacancy formation energy---4.73 eV at -0.36 volumetric strain (95 GPa pressure) to 0.07 eV at 0.33 volumetric strain (-21 GPa pressure). The pressures corresponding to volumetric strains indicated in this work are derived from the equation of state computed from a periodic calculation. Strains of the order of $10\%$ or more are not uncommon in regions exposed to a shock wave, or close to defect-cores in materials (Hoagland \textit{et al.} 1991; Kanel \textit{et al.} 1997). These results allow us to unambiguously conclude that the electronic structure of the defect-core is strongly influenced by volumetric deformations, and that this influence on the defect core-energy can not be ignored in determining the behaviour of defects governed by energetics.

\begin{figure}%
\centering%
\subfigure[]{
{\scalebox{0.9}{\includegraphics{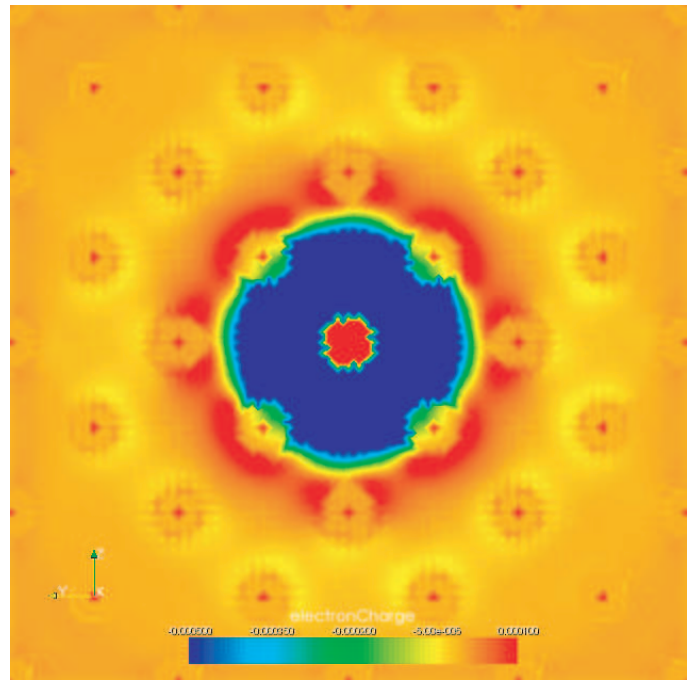}}}}\\
\subfigure[]{
{\scalebox{0.9}{\includegraphics{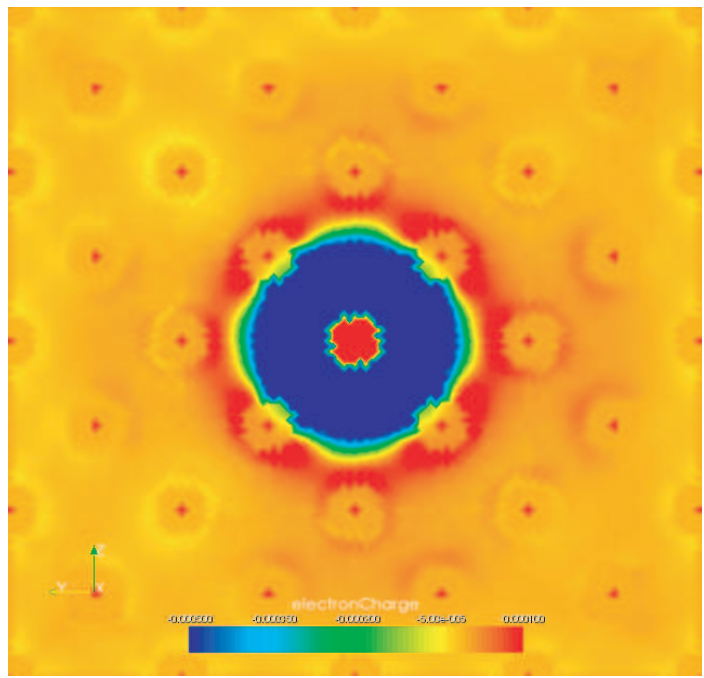}}}}
\caption{\label{vacancy_contours} Contours of electron-density perturbation around a mono-vacancy: (a) no volumetric strain; (b) 0.33 volumetric strain. These contours are plotted with a reduced range from -0.0005(blue) to 0.0001(red) to highlight the changing features with volumetric strain. The little red dots denote the positions of atoms. Adapted from Gavini (2008) for completeness of discussion.}%
\end{figure}

Electronic structure of the defect-core, initially reported in Gavini (2008), sheds further light on the influence volumetric deformations have on the core-structure. Figure \ref{vacancy_contours} shows the contours of electron-density perturbation---difference between bulk values and values in the presence of a vacancy---at the vacancy site for two cases: (a) no volumetric strain; (b) 0.33 volumetric strain. These contours are plotted with a reduced range to highlight some interesting features of the core-structure. First, as expected, the electron-density perturbation is negative around the vacancy suggesting the removal of electrons. But, there is a slight augmentation of electron-density around the first ring of atoms---this effect has been observed in previous studies (Gavini \textit{et al.} 2007b). The increase in electron-density at the centre of the vacancy is the effect of using a pseudopotential. Comparing the core electronic structure in figure \ref{vacancy_contours}~(a) and (b), some striking differences are apparent. It is evident that the contours of electron-density perturbation are moving towards the vacancy in (b) as compared to (a). This suggests shrinking of the vacancy-core with increasing volumetric deformation. Further, the nature of the electron-density perturbations are different in both cases. In (a), with no volumetric deformation, the perturbations are oscillatory, depicting the counterparts of commonly observed Friedel oscillations. Whereas in (b), the perturbations appear more diffuse. Moreover, the core appears to develop an anisotropic structure with volumetric deformation.

\begin{figure}%
\centering%
\subfigure[]{
{\scalebox{0.55}{\includegraphics{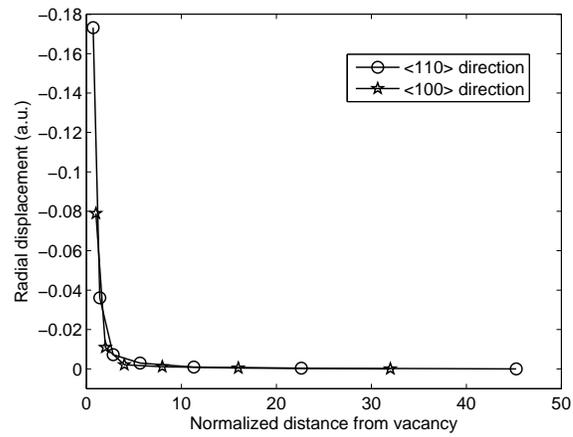}}}}\\
\subfigure[]{
{\scalebox{0.55}{\includegraphics{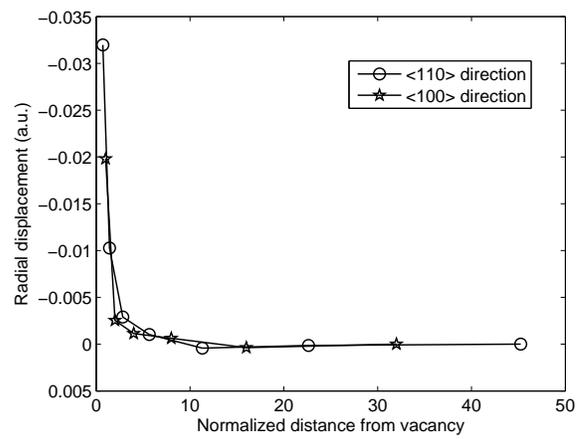}}}}
\subfigure[]{
{\scalebox{0.55}{\includegraphics{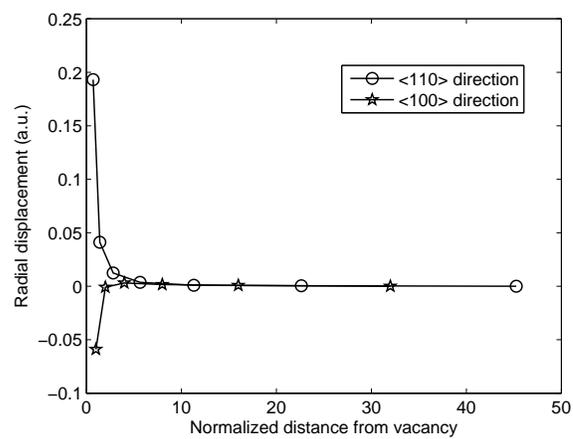}}}}
\caption{\label{vacancy_displacement} Radial displacement of atoms along $\left<110\right>$ and $\left<100\right>$ directions plotted against the distance from vacancy normalized by the lattice spacing for various imposed isotropic volumetric states of deformation. (a) -0.36 (compressive) volumetric strain; (b) undeformed crystal; (c) 0.33 (tensile) volumetric strain.}%
\end{figure}

While the electron-density contours have provided qualitative evidence on the influence of volumetric deformations on the vacancy core-structure, we provide further quantitative evidence to emphasize this dependence. We compute the displacement of atoms from their bulk positions (corresponding to the imposed macroscopic deformation) upon allowing for atomic relaxations. The computed displacement fields are predominantly radial. But, the nature and magnitude of these radial displacements along both $\left<100\right>$ and $\left<110\right>$ crystallographic directions change significantly with volumetric deformations as shown in figure \ref{vacancy_displacement}. Under $-0.36$ volumetric strain, the displacement fields are radially inward along both crystallographic directions, with a maximum displacement of 0.17 a.u. The nature of atomic relaxations are mostly radially inward for an undeformed crystal, but the maximum displacement is only 0.03 a.u. However, under $0.33 $ volumetric strain, the computed displacement fields are more complex. The displacement along $\left<110\right>$ is radially outward with a maximum displacement of 0.19 a.u. The displacement along $\left<100\right>$ follows a non-monotonic trend, being radially inward near the core, but switching to a radially outward nature away from the core. These results demonstrate that, not only the strength of the displacement field is altered by the imposed volumetric deformation, but also the nature. We note that the displacement field in the presence of a defect is a result of the atomic relaxations arising from the unbalanced forces created on the nuclei from the core electronic structure. Hence, the remarkable difference in the nature and strength of the displacement fields can be related back to the changing nature of the defect-core under volumetric deformations. The results presented thus far provide both a qualitative and quantitative confirmation of the important role of defect-cores in the energetics of vacancies that in turn determine the deformation and failure mechanisms in solids.

Next we turn to the effect of atomic relaxations on the formation energies and their dependence on the imposed volumetric deformation. From a variational stand point it is clear that atomic relaxations, which generate elastic interactions, reduce the energy of the system and consequently the formation energies. Figure \ref{vacancy_deformation}(a) shows that the relaxation energy, which is the difference between unrelaxed and relaxed states, depends on volumetric deformation. We note that the contribution to the relaxation energy comes from the defect-core, as well as, the elastic interactions that are often considered the sole contributor to the energetics in a mechanics description. The relaxation energy varies from 0.62 eV at -0.36 volumetric strain to 0.06 eV in an undeformed crystal, and is 0.41 eV at 0.33 volumetric strain. Although the variation in the relaxation energy is considerable, it is less significant compared to the variation in unrelaxed formation energy which only has contribution from the defect-core. These results, as well as similar trends for uniaxial and biaxial deformations presented subsequently, provides us with sufficient evidence to conclude that, although the elastic interactions are important, the role of the defect-core is even more significant. The computed unrelaxed formation energy of a mono-vacancy as shown in figure \ref{vacancy_deformation}(a) is positive for the range of volumetric deformations considered. However, upon relaxing the positions of atoms, which again is controlled by the defect-core, the formation energy changes sign at about 0.2 volumetric strain and is computed to be -0.34 eV at 0.33 volumetric strain. This suggests that nucleating a vacancy becomes thermodynamically favourable for large volumetric expansions. The negative formation energy at large positive volumetric deformations is a consequence of the atomic relaxations around a vacancy that release more energy than that spent in creating the vacancy. This result provides important insights into the failure mechanism of metals subjected to shock loading, which is discussed later in this section.

We now investigate the influence of volumetric deformations on  $\left<110\right>$ and $\left<100\right>$ di-vacancies, which are believed to be present in solids in large concentrations, especially at elevated temperatures (Fluss \textit{et al.} 1984; Ehrhart \textit{et al.} 1991; Hehenkamp 1994). Figure \ref{divacancy_deformation}(a) shows the relaxed di-vacancy formation energies as a function of the imposed volumetric deformation. The di-vacancy formation energies too, like the mono-vacancy formation energy, show a monotonic decrease with increasing volumetric deformation. The relaxation energies show variations with respect to volumetric strains, but are not as significant as the variations in the total formation energy. This observation is similar to the behaviour of mono-vacancies subjected to volumetric deformations, which suggests that the role of the vacancy-core appears to be more significant than the elastic interactions in the overall energetics. The di-vacancy formation energies are for the most part greater than the mono-vacancy formation energy, which is expected, as nucleating a larger defect requires more energy. However, it is interesting to note that around 0.15 volumetric strain, which corresponds to about -11 GPa pressure, both di-vacancy formation energies are lower than the mono-vacancy formation energy suggesting that it is energetically more feasible to nucleate a di-vacancy as opposed to a mono-vacancy. We attribute this behaviour to the observed larger atomic relaxations in the case of a di-vacancy as opposed to a mono-vacancy. Further, just beyond 0.15 volumetric strain, the di-vacancy formation energies, too, are found to be negative, suggesting that spontaneous nucleation of these defects is possible beyond these volumetric deformations. Figure \ref{divacancy_bind_deformation}(a) shows the influence of volumetric deformations on the di-vacancy binding energies, which is the energy released when two mono-vacancies coalesce to form a di-vacancy. The $\left<110\right>$ di-vacancy binding energy at a prescribed macroscopic strain is given by
\begin{eqnarray}
E^{bind}_{2v\left<110\right>}(\epsilon_{ij})=2E^{f}_{v}(\epsilon_{ij})-E^{f}_{2v\left<110\right>}(\epsilon_{ij}), \end{eqnarray}
and the $\left<100\right>$ di-vacancy binding energy is defined similarly. We first note that both di-vacancy binding energies are positive over the range of volumetric deformations considered. This indicates that the vacancies tend to bind under any volumetric strain. However we observe that the binding energies for $\left<110\right>$ and $\left<100\right>$ di-vacancies show different trends. While both di-vacancy binding energies show an increase with larger compressive volumetric strains, they exhibit different behaviour under the tensile regime. The results indicate that, on an average, vacancies bind more preferentially under compression as opposed to tension. Thus, although nucleating di-vacancies is energetically preferential under tensile volumetric strains, forming of di-vacancies from existing mono-vacancies appears more favourable in compression.

\subsection{Uniaxial, Biaxial, Shear deformations}

\begin{figure}%
\centering%
\subfigure[]{
{\scalebox{0.6}{\includegraphics{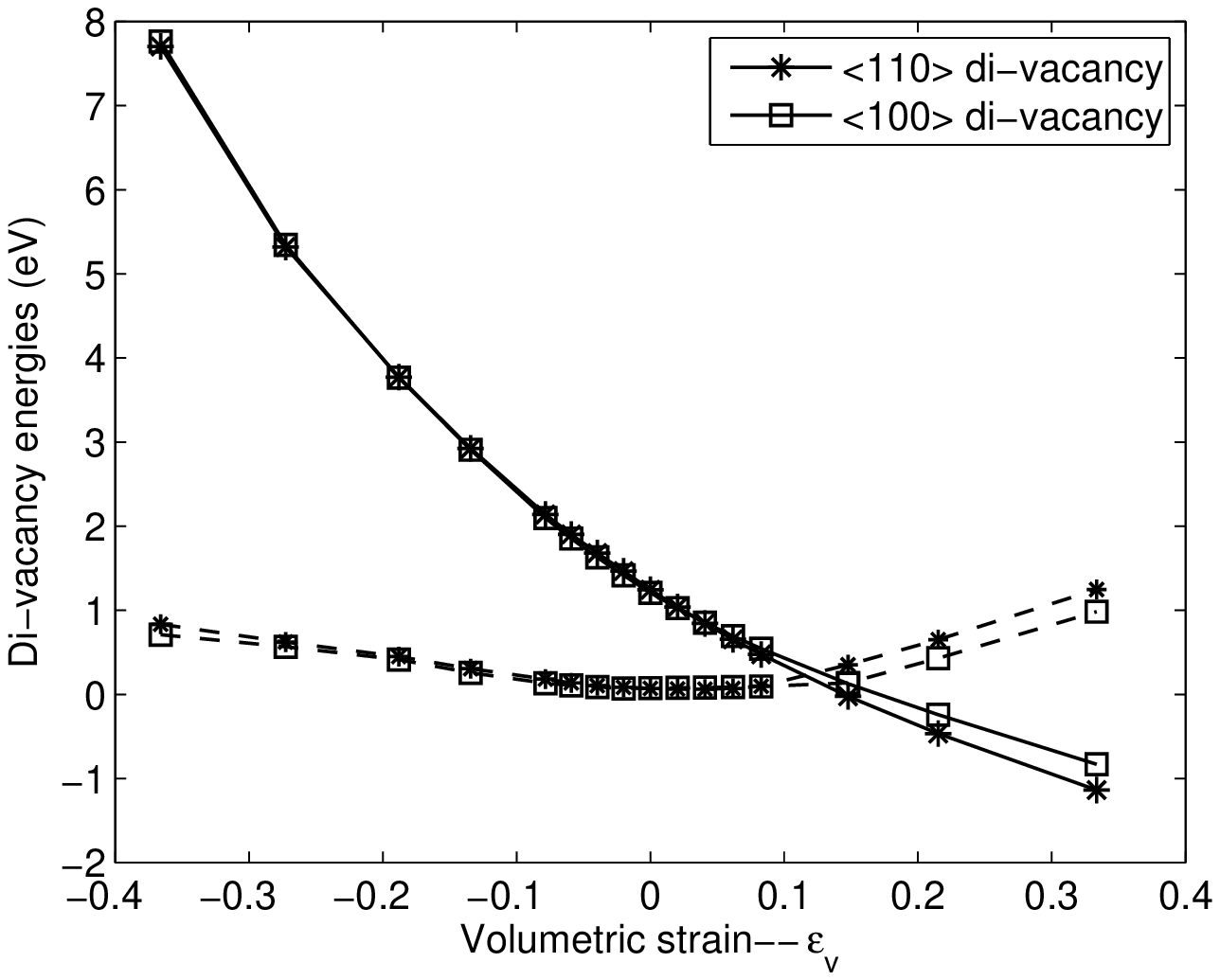}}}}%
\subfigure[]{
{\scalebox{0.6}{\includegraphics{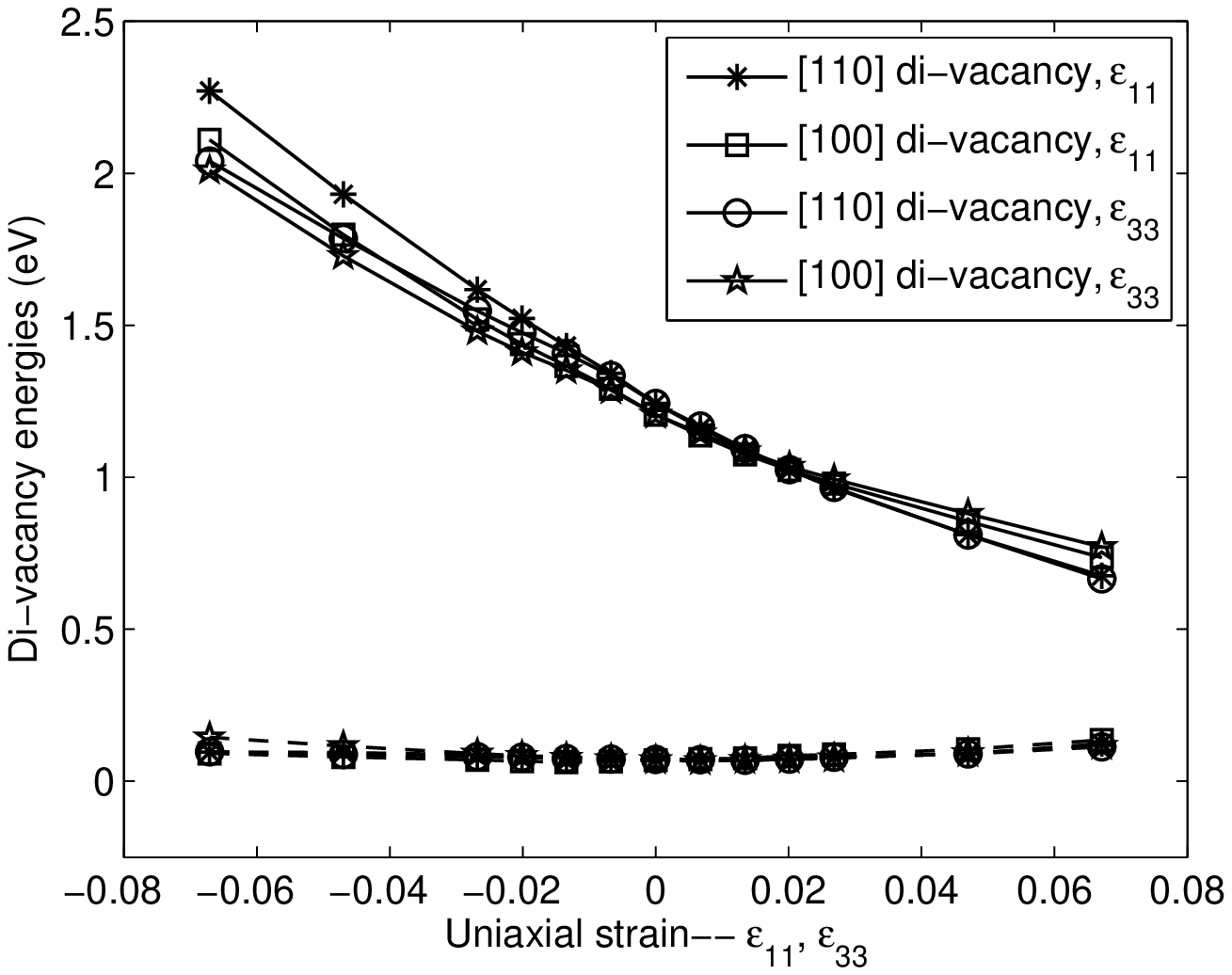}}}}\\%
\subfigure[]{
{\scalebox{0.6}{\includegraphics{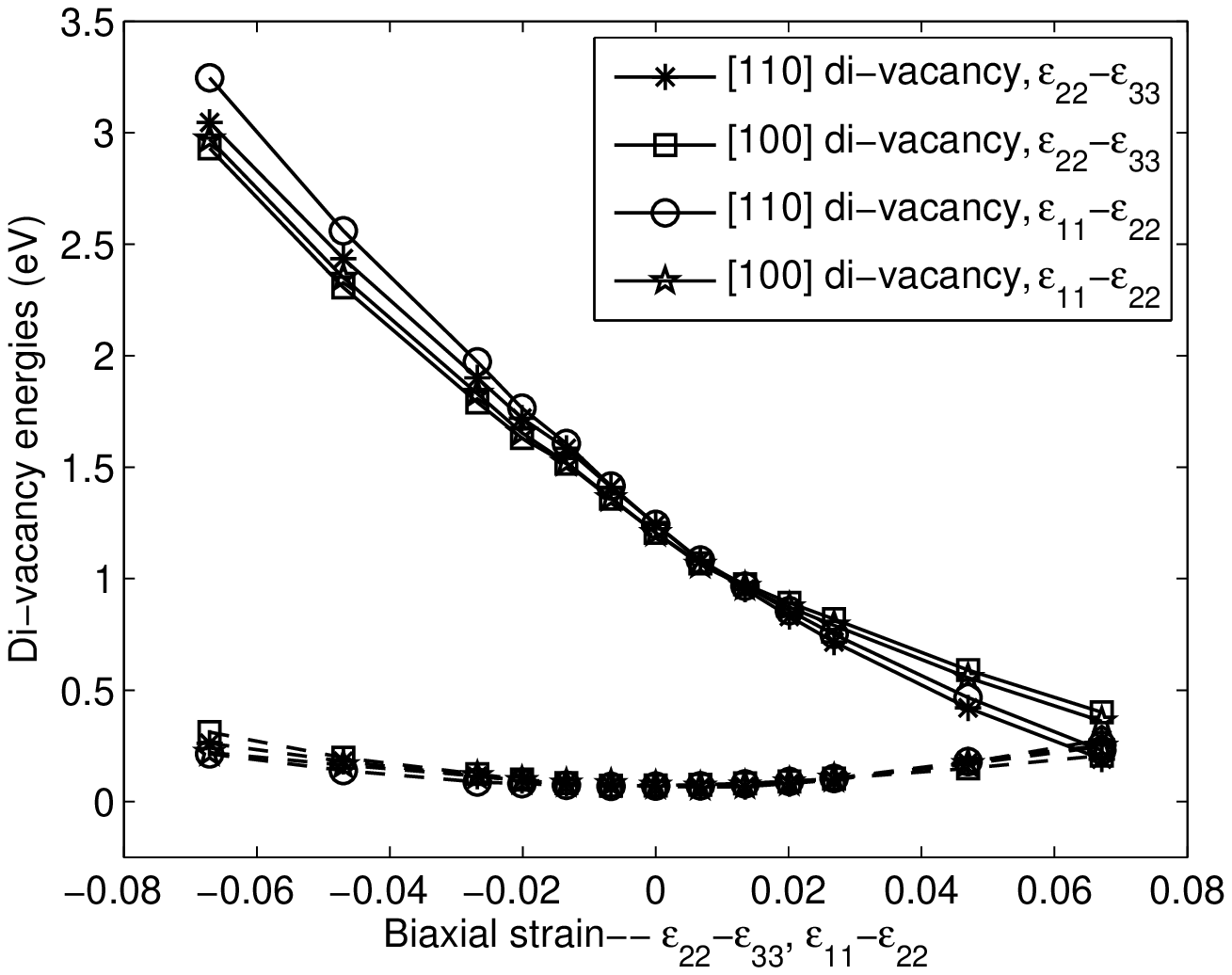}}}}%
\subfigure[]{
{\scalebox{0.6}{\includegraphics{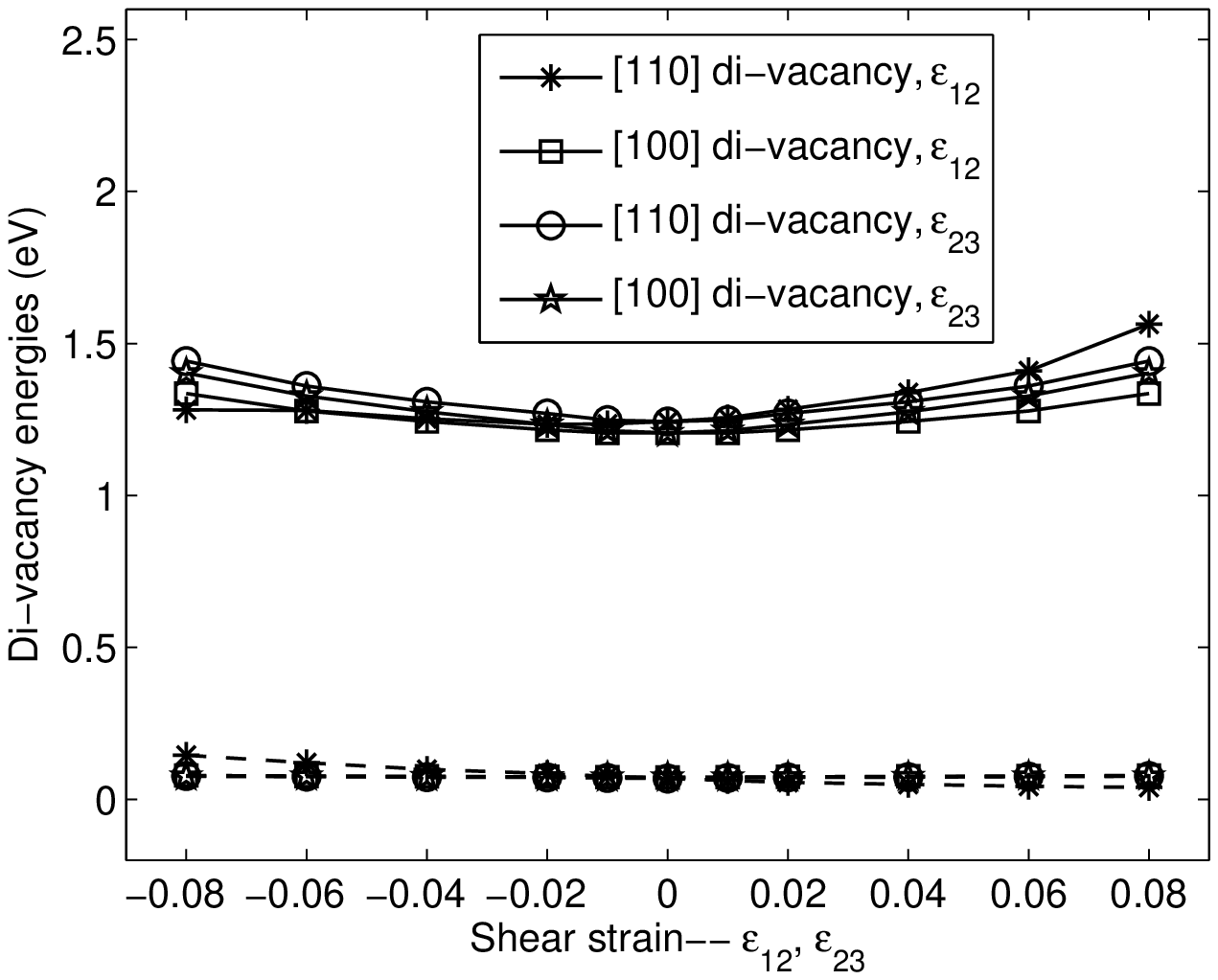}}}}%
\caption{\label{divacancy_deformation} Influence of various macroscopic deformations (strains) on the formation energies of [110] di-vacancy and [100] di-vacancy: (a) Volumetric strain; (b) Uniaxial strains; (c) Biaxial strains; (d) Shear strains. Solid lines represent relaxed formation energies, and dotted lines depict the corresponding relaxation energies.}%
\end{figure}

\begin{figure}
    \centering%
    \subfigure[]{
    {\scalebox{0.6}{\includegraphics{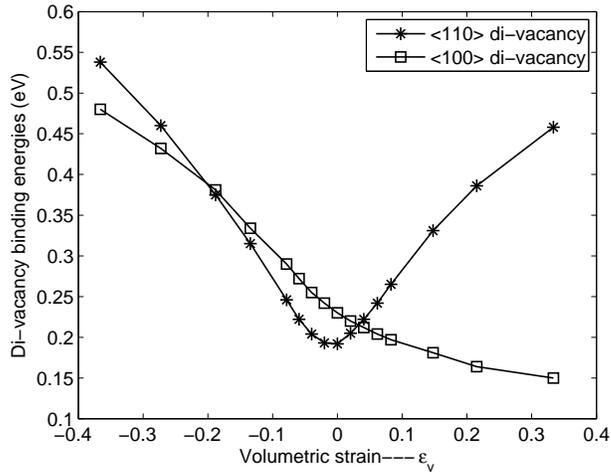}}}}%
    \subfigure[]{
    {\scalebox{0.6}{\includegraphics{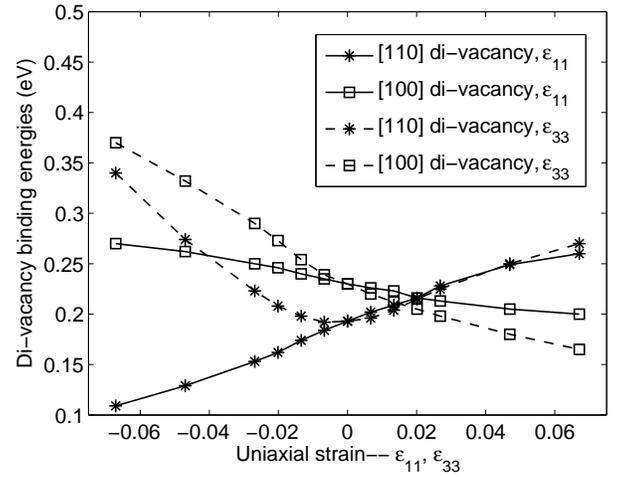}}}}\\%
    \subfigure[]{
    {\scalebox{0.6}{\includegraphics{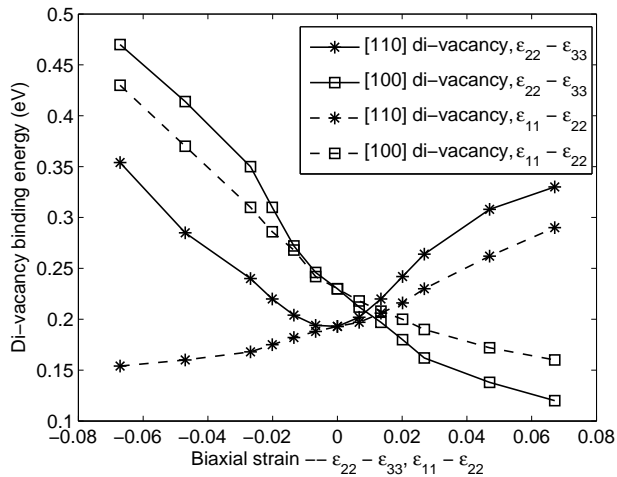}}}}%
    \subfigure[]{
    {\scalebox{0.6}{\includegraphics{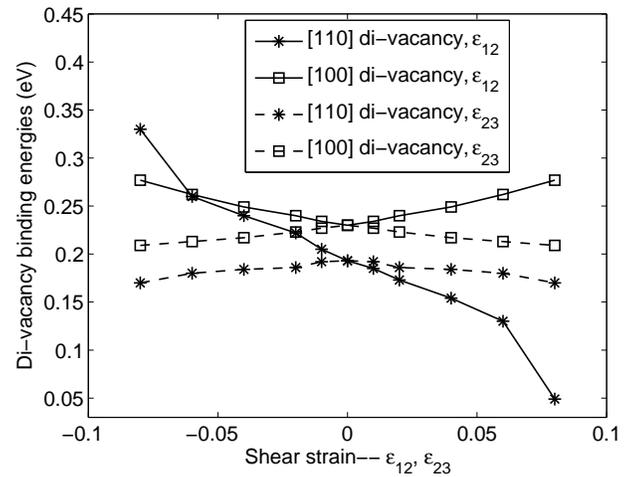}}}}%
    \caption{\label{divacancy_bind_deformation} Influence of various macroscopic deformations (strains) on the binding energies of [110] di-vacancy, and [100] di-vacancy: (a) Volumetric strain; (b) Uniaxial strains; (c) Biaxial strains; (d) Shear strains. (a), (b), (d) adapted from Gavini (2008) for completeness of the discussion.}%
\end{figure}

Having investigated the influence of isotropic volumetric deformations on the energetics of a mono-vacancy and di-vacancies, we now turn to the other commonly encountered modes of deformation---uniaxial, biaxial, and shear deformations. We first investigate the influence of various uniaxial strains on the energetics of vacancies, restricting our study to uniaxial strains along the coordinate-directions ($\left<100\right>$ directions). From symmetry, the various possible uniaxial strains, $\epsilon_{11}, \epsilon_{22}, \epsilon_{33}$, are all equivalent for a mono-vacancy. However, di-vacancies break this symmetry and $\epsilon_{11}$ and $\epsilon_{33}$ constitute the non-unique uniaxial strains possible along coordinate-directions. Figures \ref{vacancy_deformation}(b) and \ref{divacancy_deformation}(b) show the dependence of the mono-vacancy and di-vacancy formation energies on the imposed uniaxial strains. The results are qualitatively similar to the influence of volumetric deformations, representing a monotonic decrease of the formation energies, with higher energies in uniaxial compression as opposed to tension. Further, like in the case of volumetric deformations, the variation in relaxation energies of both mono-vacancy and di-vacancies are not as significant as the corresponding variation in the total formation energy. However, unlike volumetric deformations, for the range of uniaxial deformations considered, the di-vacancy formation energies are larger than the mono-vacancy formation energies, and further all energies are positive. The computed di-vacancy binding energies, shown in figure \ref{divacancy_bind_deformation}(b), exhibit variation with the imposed strain. However, no clear trends have emerged. On an average, compressive uniaxial strains seem to aid binding more than tensile states.

Figures \ref{vacancy_deformation}(c) and \ref{divacancy_deformation}(c) show the influence of biaxial states of deformation on the energetics of vacancies. As in the uniaxial deformation studies, we restrict ourselves to biaxial deformations along the coordinate-directions. This results in three possible biaxial states, of which, from symmetry, only two are non-unique for di-vacancies and one for a mono-vacancy. The computed energetics of vacancies show a similar monotonic trend as observed in the case of volumetric and uniaxial deformations, and an insignificant variation in the relaxation energies with biaxial deformations compared to the variation in the total formation energy. We note that $\left<110\right>$ di-vacancies become energetically more favourable compared to a mono-vacancy at a biaxial strain of $0.07$ or over. It is interesting that the volumetric strain corresponding to this state is about $0.15$, which is the volumetric strain where the di-vacancies were observed to become more favourable compared to mono-vacancies under isotropic volumetric deformations. As we will demonstrate later, this is not a coincidence. The computed di-vacancy binding energies shown in figure \ref{divacancy_bind_deformation}(c) are all positive suggesting vacancies attract under the range of deformations considered. Although the binding energies vary over a range of values between 0.1-0.5 eV, it appears, on an average, binding happens preferentially in compressive states of deformation, as noted previously.

Finally, we analyse the influence of shear deformations. Figures \ref{vacancy_deformation}(d) and \ref{divacancy_deformation}(d) show the dependence of the vacancy formation energies on the various shear strains which are non-unique up to symmetry. The monotonic dependence of formation energies with respect to the imposed strain, observed in all computations so far, is no longer present in shear deformations. In sharp contrast, there is no significant variation of the formation energies with respect to shear strains. The variation in the relaxation energies with shear deformations is also negligible. Further, as expected from symmetry, most formation energies exhibit an extremum with respect to shear deformations at zero strain. The computed di-vacancy binding energies also show some interesting features. Figure \ref{divacancy_bind_deformation}(d) shows that the influence of the out of plane shear ($\epsilon_{23}$) is almost negligible on binding energies of both di-vacancies. But the binding energy of [110] di-vacancy is influenced by the shear strain $\epsilon_{12}$, and unlike other shear strains this dependence is not symmetric about the origin. We note that [110] di-vacancy breaks the $C_4$---fourfold rotational---symmetry in the 1-2 plane, which results in the exhibited non-symmetric dependence on $\epsilon_{12}$ shear deformations.

\subsection{Universal role of volumetric strains}

\begin{figure}
    \centering%
    \subfigure[]{
    {\scalebox{0.6}{\includegraphics{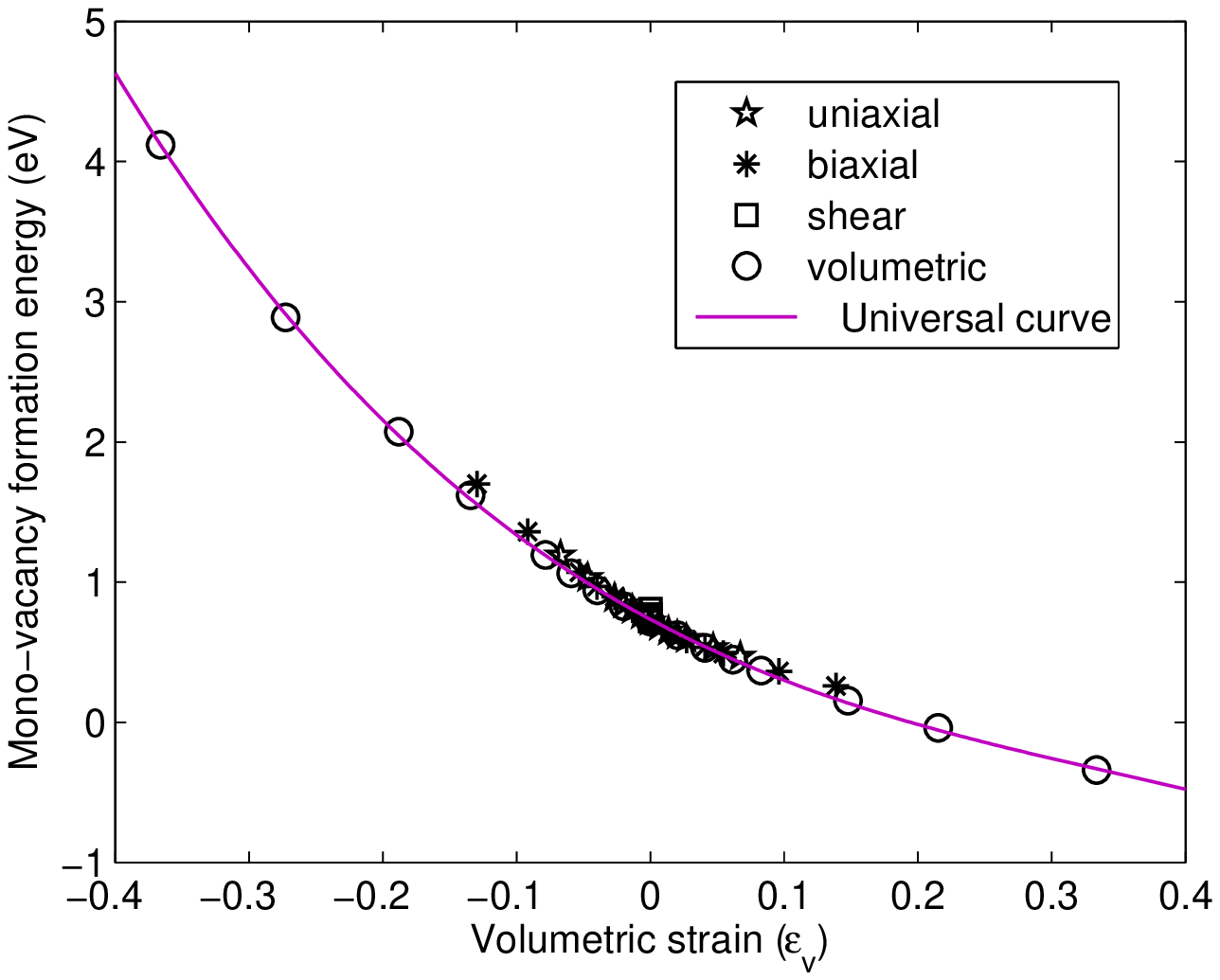}}}}%
    \subfigure[]{
    {\scalebox{0.6}{\includegraphics{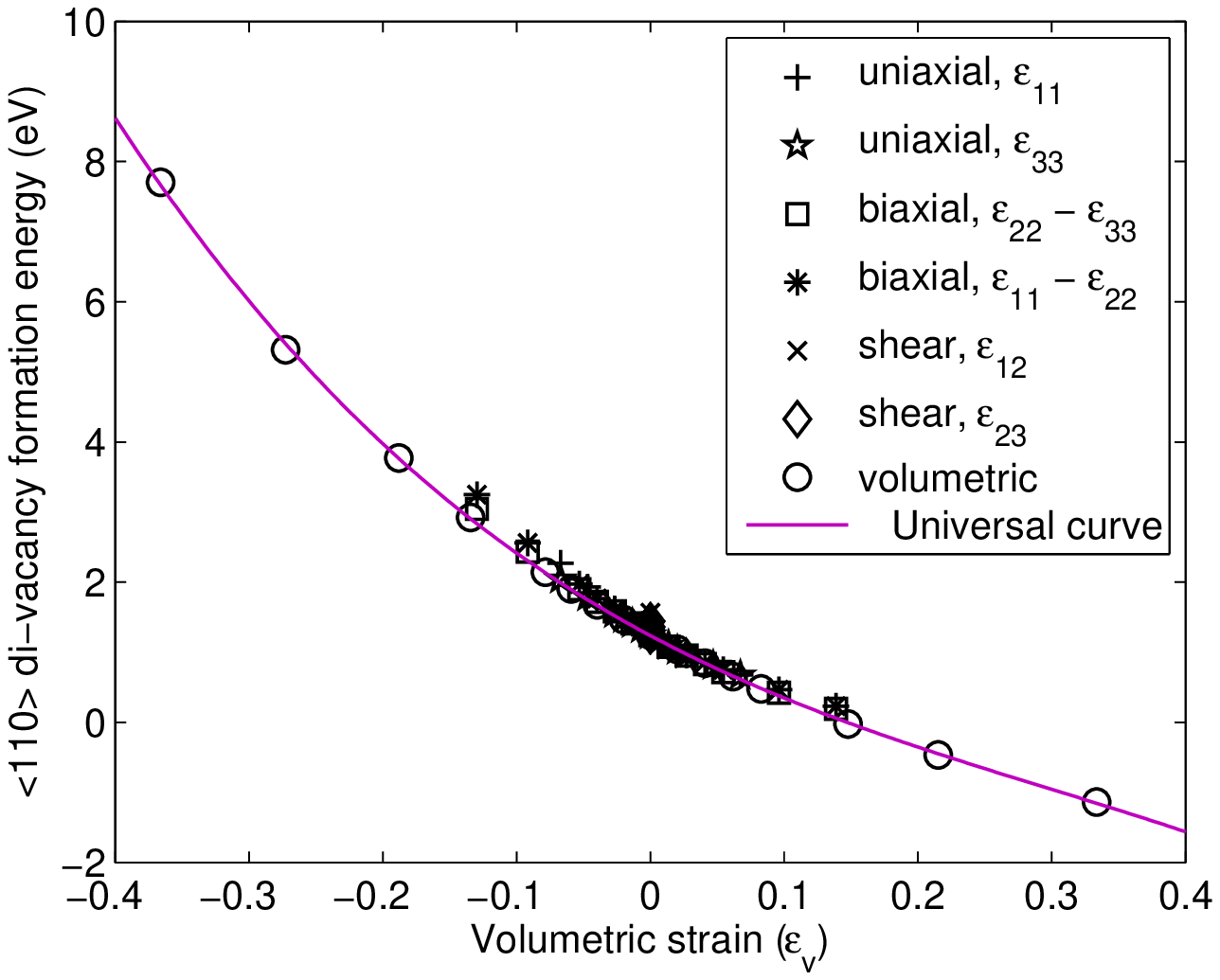}}}}\\%
    \subfigure[]{
    {\scalebox{0.6}{\includegraphics{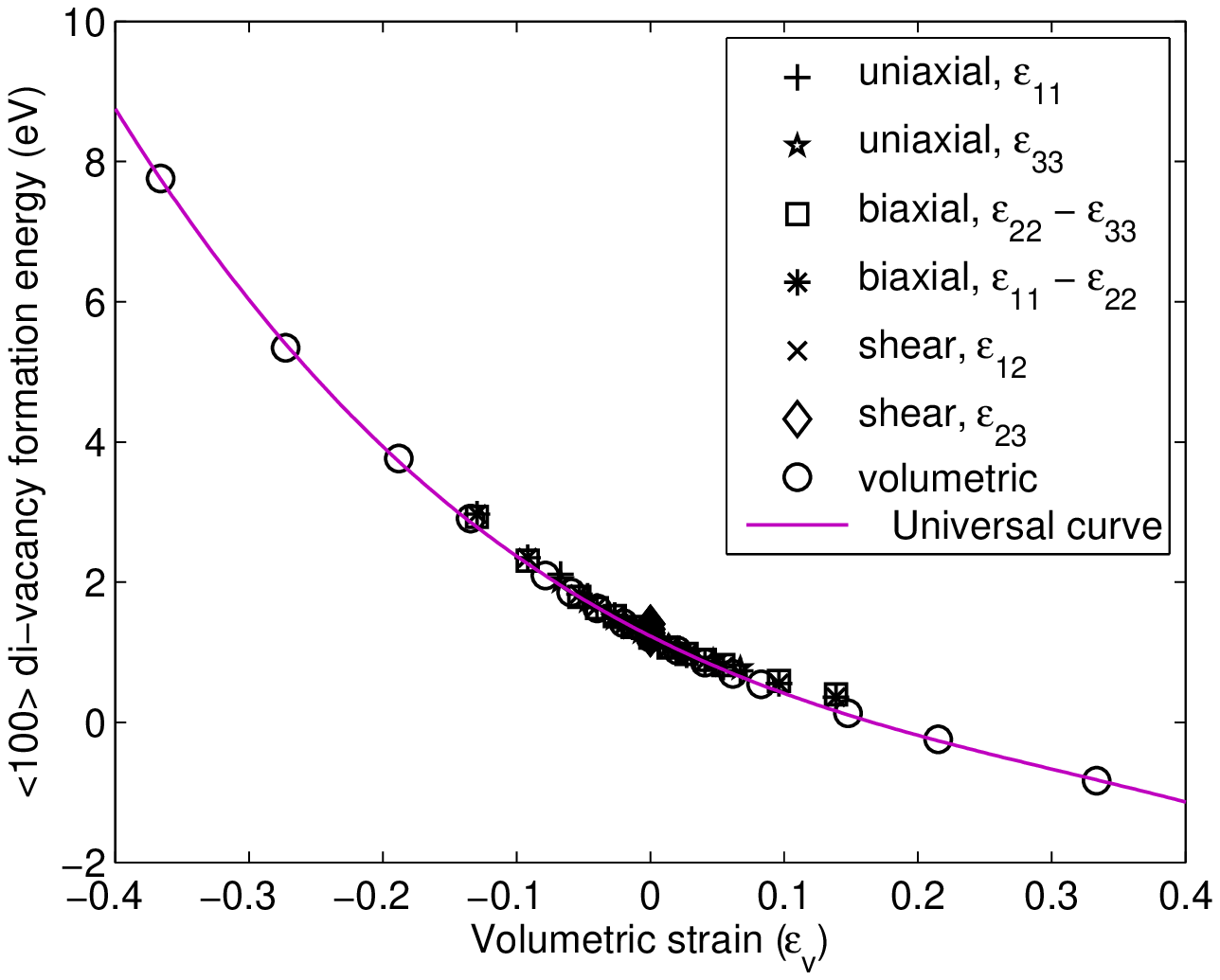}}}}%
    \caption{\label{Universal_dependence} Universal role of volumetric strains: Formation energies of mono-vacancies and di-vacancies for various deformation states---volumetric, uniaxial, biaxial, shear---are plotted against the volumetric strain associated with the deformation.}%
\end{figure}

The study on the influence of various imposed deformation states on the energetics of vacancies points to some key observations. Firstly, the monotonic dependence of the formation energies of vacancies on the imposed strain is surprisingly absent in shear deformations. We note that shear deformations do not produce volumetric strains unlike other states of deformation that were studied. Secondly, as noted previously, the volumetric strain at which di-vacancies become energetically more favourable in comparison to mono-vacancies is almost similar for isotropic volumetric deformations as well as biaxial states of deformations. Moreover, the slopes of formation energies with respect to biaxial strains is about twice as large as the corresponding slopes for uniaxial strains. We note that, for small strains, the volumetric strain corresponding to a biaxial state of deformation is twice the volumetric strain produced by a similar uniaxial state. All these observations suggest that volumetric strain associated with the imposed macroscopic deformation may be a dominant parameter governing the energetics of vacancies. To investigate this, formation energies of mono-vacancy and di-vacancies are plotted against the volumetric strain produced by the imposed macroscopic deformation. Interestingly, as shown in figure \ref{Universal_dependence}, all data points collapse onto a \emph{universal curve} with deviation in most cases less than $10\%$ of the formation energy. These results demonstrate the dominant and universal role of volumetric strains in the energetics of vacancies.

A continuum description of vacancy, which relies on elastic interactions, is often viewed as a cavity under pressure or a force dipole system (Eshelby 1951; Bullough \& Hardy 1968; Teodosiu 1982; Zhang \textit{et al.} 2006). More recently, the elastic fields produced by local lattice defects are formulated in terms of a volume tensor (Garikipati \textit{et al.} 2006), which is a convenient way to study the elastic interactions between these defects and imposed macroscopic loads/deformations. In an isotropic medium, an isotropic defect such as a vacancy produces an isotropic volume tensor, which suggests that only dilatational strain component of the imposed macroscopic deformation will contribute to elastic interaction energy between the vacancy and applied macroscopic deformation. Further, although the volume tensors associated with di-vacancies are not isotropic, the deviatoric component of the tensor is small and elastic interactions predominantly arise due to contributions from volumetric strains. Thus, the elastic interaction energy between vacancies and applied macroscopic deformations is predominantly governed by volumetric strains, which is in keeping with the observed universal role of volumetric strains in the energetics of vacancies in this study. However, we note that formation energies of vacancies have contributions from the defect-core and elastic interactions, the former being more significant than elastic energies, as discussed earlier. Since volumetric strain associated with a macroscopic deformation has been shown to be the dominant parameter influencing the formation energies of vacancies, and since elastic interactions are mainly governed by volumetric strains, we conclude that defect core-energies also are predominantly governed by volumetric strains. This key result presents a parametrization of the defect core-energies based on a single parameter---the volumetric strain produced by a deformation---for use in meso-scale models or hierarchical multi-scale schemes.

\subsection{Mechanism of spalling}

Figure \ref{Universal_dependence} shows that the energy required to nucleate mono-vacancies as well as di-vacancies monotonically decreases with increasing volumetric strains, suggesting that nucleation of these defects is increasingly favourably under volumetric expansion. Beyond a critical volumetric strain, which is smaller for di-vacancies (0.15) compared to mono-vacancies (0.2), the formation energies are computed to be negative suggesting that these defects are thermodynamically favourable at large positive volumetric deformations. This result provides a nucleation criterion for these defects and gives important physical insights into the mechanism of spalling in metals. It is observed experimentally that a large concentration of vacancies are nucleated in metals subjected to shock, which further coalesce into voids and result in material failure through spalling (Rose \& Berger 1968; Nancheva \& Saarinen 1986; Nancheva \textit{et al.} 2006; Kanel 1998). The nucleation of these defects under shock loading is often attributed to a thermally activated process due to high temperatures that are produced behind the shock front. The present results indicate that, besides the effect of temperature, the formation energies of these defects are significantly influenced by volumetric strains and this may play a crucial role in the nucleation of these defects, which has been previously ignored. Metals subjected to shock loading often undergo large volumetric deformations---both positive and negative in quick succession due to reflections from boundaries. The results presented in this article suggest that, as the expansion (tensile) wave sweeps through the material, it may leave behind a trail of vacancies and di-vacancies as the material experiences a momentary mechanical instability under these large deformations and nucleates vacancies. Although the nucleation of these defects is thermodynamically favourably under large deformations, as noted earlier, their coalescence is preferred under compressive strains. It is likely that, under the compression wave which follows the expansion wave, the nucleated vacancies coalesce to form larger voids. As this cycle repeats, we believe large voids nucleate and the material eventually fails due to spalling.

\begin{figure}%
\centering
{\scalebox{0.6}{\includegraphics{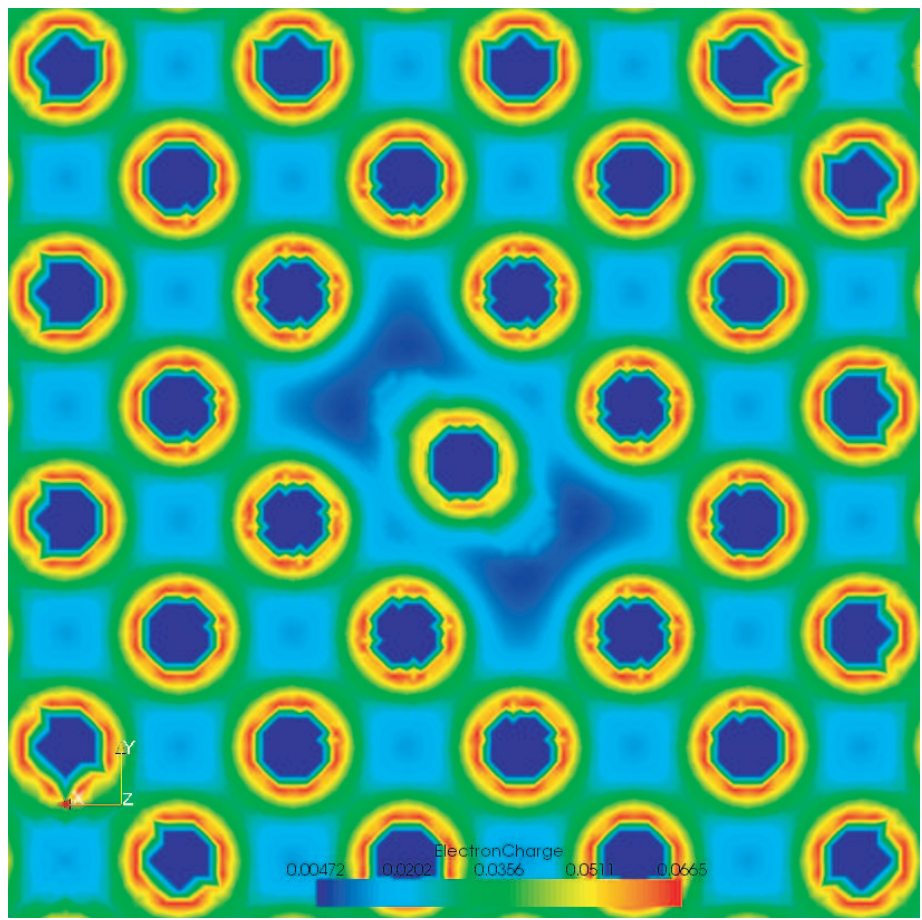}}}
\caption{\label{vacancy_migration_contours} Electron-density contours along (001) plane of a $[1\overline{1}0]$ vacancy migration at the saddle point.}%
\end{figure}

\begin{figure}%
\centering
{\scalebox{1.0}{\includegraphics{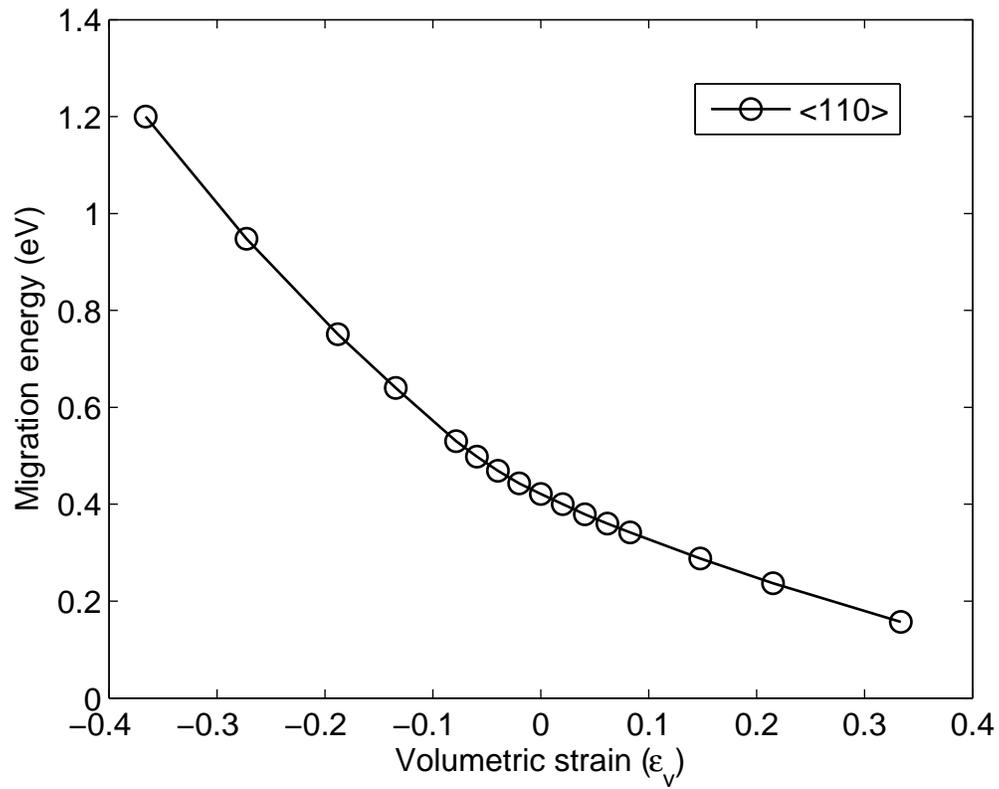}}}
\caption{\label{vacancy_migration} Dependence of $\left<110\right>$ vacancy migration energy on volumetric strains.}%
\end{figure}

Though the present investigation provides new insights into the fundamental mechanism governing spalling in materials, the time-scales involved in the failure process are governed by the kinetics of these defects. It can be expected that the kinetics, which are governed by the migration energies, are also influenced by volumetric strains. To this end, we study the dependence of migration energies on volumetric strains. We restrict our study to the migration of a single vacancy, as the migration energies of larger defects are considerably larger and a single vacancy migration is the most probable. We investigate the energies associated with two hops: (i) along $\left<100\right>$---second nearest neighbour hop; (ii) along $\left<110\right>$---nearest neighbour hop. We compute the migration energy of a hop as the difference in the relaxed energies when the migrating atom is at the saddle-point and at the lattice site. We assume the saddle-point for these hops exists midway along the shortest distance path of the hop (high-symmetry point), and later verify this by perturbing the position of the atom (though arbitrarily) along the high-symmetry plane and away from it. Figure \ref{vacancy_migration_contours} shows the contours of electron-density when the atom is at the high-symmetry point. Though more accurate methods of locating the saddle-point exist, like the nudged elastic band (Henkelman \& J\'{o}nsson 2000), which perform a global search and locate the saddle-point with least energy, they have currently not been integrated with the quasi-continuum approach. The migration energy of a $\left<100\right>$ hop in an undeformed crystal is computed as 1.3 eV as opposed to 0.42 eV for $\left<110\right>$ hop. This suggests that a $\left<100\right>$ hop is highly unlikely, and migration along $\left<100\right>$ directions happens through two hops along $\left<110\right>$ directions. Thus, we restrict our investigation to the influence of volumetric strains on $\left<110\right>$ migration energies. Figure \ref{vacancy_migration} shows the variation of the $\left<110\right>$ migration energy with volumetric strain. The migration energy exhibits a monotonic decrease with increasing volumetric strain. This suggests that vacancies tend to migrate far more easily under volumetric expansions as opposed to compression. In the context of spall failure in materials, nucleation and kinetics of vacancies are most active under the expansion wave, whereas vacancy coalescence, on an average, appears favourable under compression.

\section{Conclusions}\label{Conclusions}

In this work, we investigated the effect of macroscopic deformations on the energetics of vacancies in aluminum. By suppressing the atomic relaxations, we first studied the energetics and nature of the defect-core under varying volumetric strains. We found that vacancy core-energy is strongly influenced by volumetric deformations, and that this dependence is more significant than the elastic interactions arising out of atomic relaxations. Thus, the defect-core appears to play a significant role in the energetics of vacancies, which has been previously ignored in continuum descriptions that formulate the energetics of defects solely based on elastic interactions. By observing the electron-density contours and the nature of displacement fields, we believe that the electronic structure is significantly altered by the local state of deformation at the core. The core appeared to shrink with increasing volumetric strain and also developed an anisotropic nature. The changing nature of the core also translated into varying trends in the displacement fields. While the computed displacement field was radially inward for volumetric compression and in an undeformed crystal, it was predominantly radially outward for volumetric expansions. The energetics and insights into the core-structure emphasize the crucial role defect-cores play in determining the behaviour of vacancies.

We further studied the influence of other states of deformation---uniaxial, biaxial, and shear deformations---on the energetics of vacancies. It appeared that volumetric strain associated with a state of deformation is a dominant parameter influencing the energetics of vacancies. Upon plotting the formation energies of vacancies computed for various states of deformation against the volumetric strain associated with the deformation, all data points collapse onto a master curve, suggesting that the role of volumetric strains in the energetics of vacancies may be universal. Considering that elastic fields generated by vacancies mainly interact with volumetric strains, we conclude that defect-core energies, too, are significantly influenced by the volumetric strain component of a deformation. This provides a one-variable parametrization of the dependence of vacancy core-energy on the strain tensor.

The dependence of vacancy formation energies on volumetric deformations provides new insights into the phenomena of spalling in metals exposed to shock waves. We note that formation energies of vacancies become negative for large positive volumetric strains, which are quite common in shock conditions, suggesting that vacancies tend to nucleate spontaneously. This is a result of the instability in the material created at very large strains, where, by nucleating vacancies the system achieves a lower energy by local atomic relaxations. The present study indicates that changes in formation energies are significant when a material is subjected to large deformations and have to be accounted for in determining the appropriate thermodynamic quantities. Unlike the common belief that vacancy nucleation during shock loading is a thermally activated process, the present study suggests that this nucleation may be a mechanically activated process and reemphasizes the important role of the defect-core in determining the behaviour of vacancies. Though clear trends have emerged in the dependence of formation energies on volumetric strains, the di-vacancy binding energies did not show any conceivable pattern. However, on an average, it appeared that vacancies tend to bind preferentially under compressive deformations as opposed to tensile states. The $\left<110\right>$ migration energy also showed a significant dependence on volumetric strains, indicating the important influence of volumetric deformations on the nucleation and kinetics of vacancies.

The role of a defect-core in energetics has been demonstrated through studies on vacancies. Like vacancies, the dependence of core-energy on the local state of deformation at the core is likely in the case of other defects like dislocations, and presents itself as a worthwhile direction for future work. Further, we have restricted our investigations to formation energies by controlling the macroscopic deformation. Though formation energies have provided crucial insights into the behaviour of defects and are interesting in their own respect, another thermodynamic variable of considerable interest is the formation enthalpy when macroscopic stress state is controlled. This presents another important avenue for future investigations. On the end of method developments, formulating the quasi-continuum reduction for Kohn-Sham density functional theory presents a significant challenge owing to the delocalized nature of the wavefunctions and the global orthogonality constraints on them. Though challenging, it is one of the outstanding problems that will enable large scale electronic structure calculations on a wide range of materials systems that are not adequately described by orbital-free approaches.

\begin{acknowledgements}
The author gratefully acknowledges the support of the Air Force Office of Scientific Research under Grant no. FA9550-09-1-0240.
\end{acknowledgements}


\section{References}
\begin{thedemobiblio}{99}
\item
Acharya, M. \& Bassani, L. 2000 Lattice incompatibility and a gradient theory of crystal plasticity. \emph{J. Mech. Phys. Solids} \textbf{48}, 1565--1595.
\item
Blanc, X., Le Bris, C., \& Lions, P. L. 2002 From molecular models to continuum mechanics. \emph{Arch. Rational Mech. Anal.} \textbf{164}, 341--381.
\item
Brenner, S. C. \& Scott, L. R. 2002 \emph{The mathematical theory of finite element methods}. New York: Springer-Verlag.
\item
Bullough, R. \& Hardy, J. R. 1968 The strain field interaction between vacancies in copper and aluminium. \emph{Phil. Mag.} \textbf{17}, 833--842.
\item
Carling, K., Wahnstr\"om, G., Mattsson, T. R., Mattsson, A. E., Sandberg, N. \& Grimvall, G. (2000) Vacancies in metals: From first-principles calculations to experimental data. \emph{Phys. Rev. Lett.} \textbf{85}, 3862--3865.
\item
Ceperley, D. M. \& Alder, B. J. 1980 Ground state of the electron gas by a stochastic method. \emph{Phys. Rev. Lett.} \textbf{45}, 566--569.
\item
Chetty, N., Weinert, M., Rahman, T. S. \& Davenport, J. W. 1995 Vacancies and impurities in aluminum and magnesium. \emph{Phys. Rev. B} \textbf{52}, 6313--6326.
\item
Choly, N., \& Kaxiras, E. 2002 Kinetic energy density functionals for non-periodic systems. \emph{Solid State Communications} \textbf{5}, 281--286.
\item
Dal Maso, G. 1993 \emph{An introduction to $\Gamma$-convergence}. Boston: Birkh\"auser.
\item
Ehrhart, P., Jung, P., Schultz, H., \& Ullmaier, H. 1991 \emph{Atomic defects in metal,} Landolt-B$\ddot{o}$rnstein, New Series, Group 3, Vol. 25. Berlin: Springer-Verlag.
\item
Eshelby, J. D. 1951 The force on an elastic singularity. Phil. Trans. R. Soc. Lond. A \textbf{244}, 87--112.
\item
Eshelby, J. D. 1956 The continuum theory of lattice defects. In \emph{Solid State Physics, Advances in Research and
Applications } (eds F. Seitz and D. Turnbull), pp.~79-144, Vol. III. New York: Academic Press.
\item
Fluss, M. J., Berko, S., Chakraborty, B., Hoffmann, K. R., Lippel, P., \& Siegel, R. W. 1984 Positron annihilation spectroscopy of the equilibrium vacancy ensemble in aluminium. \emph{J. Phys. F : Met. Phys.} \textbf{14}, 2831--2854.
\item
Finnis, M. 2003 \emph{Interatomic forces in condensed matter}. Oxford: Oxford University Press.
\item
Garikipati, K., Falk, M., Bouville, M., Puchala, B. \& Narayanan, H. 2006 The continuum elastic and atomistic viewpoints on the formation volume and strain energy of a point defect. \emph{J. Mech. Phys. Solids} \textbf{54}, 1929--1951.
\item
Gavini, V., Knap, J., Bhattacharya, K. \& Ortiz, M. 2007a
Non-periodic finite-element formulation of orbital-free
density-functional thoery. \emph{J. Mech. Phys. Solids} \textbf{55}, 669--696.
\item
Gavini, V., Bhattacharya, K. \& Ortiz, M. 2007b Quasi-continuum
orbital-free density-functional theory: A route to multi-million
atom non-periodic DFT calculation. \emph{J. Mech. Phys. Solids} \textbf{55}, 697--718.
\item
Gavini, V., Bhattacharya, K. \& Ortiz, M. 2007c Vacancy clustering and prismatic dislocation loop formation in aluminum. \emph{Phys. Rev. B} \textbf{76}, 180101--180105.
\item
Gavini, V. 2008 Role of macroscopic deformations in energetics of vacancies in aluminum. \emph{Phys. Rev. Lett.} \textbf{101}, 205503--205507.
\item
Gillan, M. J. 1989 Calculation of the vacancy formation energy in aluminium. \emph{J. Phys.: Condens. Matter} \textbf{1}, 689--711.
\item
Goodwin, L., Needs, R. J., \& Heine, V. 1990 A pseudopotential total energy study of impurity promoted intergranular embrittlement. \emph{J. Phys. Condens. Matter} \textbf{2}, 351--365.
\item
Hehenkamp, T., 1994 Absolute vacancy concentrations in noble metals and some of their alloys. \emph{J. Phys. Chem. Solids} \textbf{55}, 907--915.
\item
Henkelman, G., \& J\'{o}nsson, H. 2000 Improved tangent estimate in the nudged elastic band method for finding minimum energy paths and saddle points. \emph{J. Chem. Phys.} \textbf{113}, 9978--9985.
\item
Hoagland, R. G., Daw, M. S., \& Hirth, J. P. 1991 Some aspects of forces and fields in atomic models of crack tips. \emph{J. Mater. Res.} \textbf{6}, 2565--2577.
\item
Kanel, G. I., Razorenov, S. V., Bogatch, A., Utkin, A. V., \& Grady, D. E. 1997 Simulation of spall fracture of aluminum and magnesium over a wide range of load duration and temperature. \emph{Int. J. Impact Eng.} \textbf{20}, 467--478.
\item
Kanel, G. 1998 Some new data on deformation and fracture of solids under shock-wave loading. \emph{J. Mech. Phys. Solids} \textbf{46}, 1869--1886.
\item
Knap, J. \& Ortiz, M. 2001 An analysis of the quasicontinuum method. \emph{J. Mech. Phys. Solids} \textbf{49}, 1899--1923.
\item
Lu, G. \& Kaxiras, E. 2002 Can vacancies lubricate dislocation motion in aluminum? \emph{Phys. Rev. Lett.} \textbf{89}, 105501--105505.
\item
Lu, G. \& Kaxiras, E. 2005 Hydrogen embrittlement of aluminum: The crucial role of vacancies. \emph{Phys. Rev. Lett.} \textbf{94}, 155501--155505.
\item
Martin, M. 2004 \emph{Electronic structure, basic theory and practical
methods}. Cambridge: Cambridge University Press.
\item
Meyers, M. A. \& Aimone, C. T. 1983 Dynamic fracture (spalling) of metals. \emph{Progress in Materials Science} \textbf{28}, 1--96.
\item
Nancheva, N. M. \& Saarinen, K. 1986 Positron lifetime studies of shock loaded nickel. \emph{Scripta Metallurgica} \textbf{20}, 1085--1088.
\item
Nancheva, N. M., Saarinen, K. \& Popov, G. S. 2006 Positron annihilation in shock loaded titanium and titanium alloy BT14. \emph{Physica Status Solidi (a)} \textbf{95}, 531--536.
\item
Parr, R. G. \& Yang, W. 1989 \emph{Density-functional theory of atoms and molecules}. Oxford: Oxford University Press.
\item
Pask, J. E., Klein, B. M., Fong, C. Y. \& Sterne, P. A., 1999 Real-space local polynomial
basis for solid-state electronic-structure calculations: A finite-element approach. \emph{Phys. Rev. B} \textbf{59}, 12352–12358.
\item
Pask, J. E. \& Sterne, P. A., 2005 Finite element methods in ab initio electronic structure
calculations. \emph{Modelling Simul. Mater. Sci. Eng.} \textbf{13}, R71–R96.
\item
Perdew, J. P. \& Zunger, A. 1981 Self-interaction correction to density-functional approximations for many-electron systems. \emph{Phys. Rev. B} \textbf{23}, 5048--5079.
\item
Phillips, R. 2001 \emph{Crystals, defects, and microstructure: Modelling across scales}. Cambridge: Cambridge University Press.
\item
Rose, M. F. \& Berger, T. L. 1968 Shock deformation of polycrystalline aluminium. \emph{Phil. Mag.} \textbf{17}, 1121--1133.
\item
Smargiassi, E. \& Madden, P. A. 1994 Orbital-free kinetic-energy functionals for first-principles molecular dynamics. \emph{Phys. Rev. B} \textbf{49}, 5220--5226.
\item
Somoza, A., Macchi, C. E., Lumley, R. N., Polmear, I. J., Dupasquier, A. \& Ferragut, R. 2007 Role of vacancies during creep and secondary precipitation in an underaged Al-Cu-Mg-Ag. \emph{Physica Status Solidi (c)} \textbf{4}, 3473--3476.
\item
Tadmor, E. B., Ortiz, M. \& Phillips, R. 1996 Quasicontinuum analysis of defects in solids. \emph{Philos. Mag. A} \textbf{73}, 1529--1563.
\item
Teodosiu, C. 1982 \emph{Elastic Models of Crystal Defects}. Berlin: Springer.
\item
Triftsh\"auser, W. 1975 Positron trapping in solid and liquid metals. \emph{Phys. Rev. B} \textbf{12}, 4634--4639.
\item
Turner, D. E., Zhu, Z. Z., Chan, C. T., \& Ho, K. M. 1997 Energetics of vacancy and substitutional impurities in aluminum bulk and clusters. \emph{Phys. Rev. B} \textbf{55}, 13842--13852.
\item
Uesugi, T., Kohyama, M. \& Higashi, K. 2003 Ab initio study on divacancy binding energies in aluminum and magnesium. \emph{Phys. Rev. B} \textbf{68}, 184103--184108.
\item
Wang, Y. A., Govind, N. \& Carter, E. A. 1998 Orbital-free kinetic-energy functionals for the nearly free electron gas. \emph{Phys. Rev. B} \textbf{58}, 13465--13471.
\item
Wang, Y. A., Govind, N. \& Carter, E. A. 1999 Orbital-free kinetic-energy density functionals with a density-dependent kernel. \emph{Phys. Rev. B} \textbf{60}, 16350--16358.
\item
Wang, L. \& Teter, M. P., 1992 Kinetic energy functional of electron density. \emph{Phys. Rev. B}, \textbf{45}, 13196-13220.
\item
Zhang, X., Jiao, K., Sharma, P., \& Yakobson, B. I. 2006 An atomistic and non-classical continuum field theoretic perspective of elastic interactions between defects (force dipoles) of various symmetries and application to graphene. \emph{J. Mech. Phys. Solids} \textbf{54}, 2304--2329.
\end{thedemobiblio}

\newpage

\newpage

\newpage

\newpage

\newpage

\newpage

\newpage

\newpage

\newpage

%
%

\end{document}